\documentclass[fleqn,usenatbib, useAMS]{mnras}

\usepackage{newtxtext,newtxmath}

\usepackage[T1]{fontenc}
\usepackage{ae,aecompl}
\usepackage{graphicx}	
\usepackage{times}

\title[The mass of the Milky Way]{The mass of the Milky Way out to 100 kpc using halo stars}

\author[A. J. Deason et al.]{
\parbox{\textwidth}{
  \Large
  Alis J. Deason$^{1,2}$\thanks{alis.j.deason@durham.ac.uk},
  Denis Erkal$^{3}$,
  Vasily Belokurov$^{4}$,
  Azadeh Fattahi$^{1}$,
  Facundo A. G\'{o}mez$^{5,6}$,
  \newline
  Robert J. J. Grand$^{7}$,
  R\"udiger Pakmor$^{7}$,
  Xiang-Xiang Xue$^{8}$,
  Chao Liu$^{9}$,
  Chengqun Yang$^{10}$,
  Lan Zhang$^{8}$
  \newline
  and Gang Zhao$^{8}$
}
\vspace{0.25cm}
\\
$^{1}$Institute for Computational Cosmology, Department of Physics, University of Durham, South Road, Durham DH1 3LE, UK\\
$^{2}$Centre for Extragalactic Astronomy, Department of Physics, University of Durham, South Road, Durham DH1 3LE, UK\\
$^{3}$Department of Physics, University of Surrey, Guildford GU2 7XH, UK \\
$^{4}$Institute of Astronomy, Madingley Rd, Cambridge, CB3 0HA \\
$^{5}$Instituto de Investigaci\'on Multidisciplinar en Ciencia y Tecnolog\'ia, Universidad de La Serena, Ra\'ul Bitr\'an 1305, La Serena, Chile\\
$^{6}$Departamento de Astronom\'ia, Universidad de La Serena, Av. Juan Cisternas 1200 Norte, La Serena, Chile\\
$^{7}$Max-Planck-Institut f\"{u}r Astrophysik, Karl-Schwarzschild-Str. 1, 85748 Garching, Germany\\
$^{8}$CAS Key Laboratory of Optical Astronomy, National Astronomical Observatories, Beijing 100101, China\\
$^{9}$Key lab of Space Astronomy and Technology, National Astronomical Observatories, Beijing 100101, China\\
$^{10}$Shanghai Astronomical Observatory, 80 Nandan Road, Shanghai 200030, China
}

\date{Accepted XXX. Received YYY; in original form ZZZ}

\pubyear{2020}

\begin{document}
\label{firstpage}
\pagerange{\pageref{firstpage}--\pageref{lastpage}}
\maketitle

\begin{abstract}
 We use a distribution function analysis to estimate the mass of the Milky Way out to 100 kpc using a large sample of halo stars. These stars are compiled from the literature, and the vast majority ($\sim \! 98\%$) have 6D phase-space information. We pay particular attention to systematic effects, such as the dynamical influence of the Large Magellanic Cloud (LMC), and the effect of unrelaxed substructure. The LMC biases the (pre-LMC infall) halo mass estimates towards higher values, while realistic stellar halos from cosmological simulations tend to underestimate the true halo mass. After applying our method to the Milky Way data we find a mass within 100 kpc of $M(< 100 \mathrm{kpc}) = 6.07 \pm 0.29 \mathrm{(stat.)} \pm 1.21 \mathrm{(sys.)}  \times 10^{11}$M$_\odot$.  For this estimate, we have approximately corrected for the reflex motion induced by the LMC using the Erkal et al. model, which assumes a rigid potential for the LMC and MW. Furthermore, stars that likely belong to the Sagittarius stream are removed, and we include a 5\% systematic bias, and a 20\% systematic uncertainty based on our tests with cosmological simulations. Assuming the mass-concentration relation for Navarro-Frenk-White haloes, our mass estimate favours a total (pre-LMC infall) Milky Way mass of $M_{\rm 200c} = 1.01 \pm 0.24  \times 10^{12}$M$_\odot$, or (post-LMC infall) mass of $M_{\rm 200c} = 1.16 \pm 0.24  \times 10^{12}$ M$_\odot$ when a $1.5 \times 10^{11}$M$_\odot$ mass of a rigid LMC is included.
 
\end{abstract}


\section{Introduction}
The total mass of the Milky Way has been an historically difficult parameter to pin down (see recent reviews by \citealt{bland-hawthorn16, wang19}). Despite decades of measurements, there remains an undercurrent of elusiveness surrounding \textit{``the mass of the Milky Way''}. However, the continued eagerness to provide an accurate measure is perhaps unsurprising --- the mass of a halo is arguably its most important characteristic. For example, almost every property of a galaxy is dependent on its halo mass, and thus this key property is essential to place our ``benchmark'' Milky Way galaxy in context within the general galaxy population. In addition, the host halo mass is inherently linked to its subhalo population, so most of the apparent small scale discrepancies with the $\Lambda$CDM model \citep[e.g.][]{moore99, boylan-kolchin11} are strongly dependent on the Milky Way mass \citep[see e.g.][]{wang12}. Moreover, tests of alternative dark matter candidates critically depend on the total mass of the Milky Way, particularly for astrophysical tests \citep[e.g.][]{kennedy14, lovell14}. 

The uncertainty has stemmed from two major shortcomings: (1) a lack of luminous tracers with full 6D phase-space information out to the viral radius of the Galaxy, and (2) an underestimated, or unquantified, systematic uncertainty in the mass estimate. However, there has been significant progress since the first astrometric data release from the \textit{Gaia} satellite \citep{gdr2}. This game-changing mission for Milky Way science provided the much needed tangential velocity components for significant numbers of halo stars, globular clusters and satellite galaxies. Indeed, there are encouraging signs that we are converging to a total mass of just over $1 \times 10^{12}$M$_\odot$ \citep[e.g.][]{callingham19, deason19, eadie19, grand19, vasiliev19, watkins19, cautun20, li20}. However, mass estimates at very large distances (i.e. beyond 50 kpc), are dominated by measures using the kinematics of satellite galaxies, which probe out to the virial radius of the Galaxy \citep[e.g.][]{patel18, callingham19, li20}.  It is well-known that the dwarf satellites of the Milky Way have a peculiar planar alignment \citep[see e.g.][]{metz07}, and, without independent measures at these large distances, there remains uncertainty over whether or not the satellites are biased kinematic tracers of the halo.  

Arguably the most promising tracers at large radii are the halo stars. They are significantly more numerous than the satellite galaxies and globular clusters, and are predicted to reach out to the virial radius of the Galaxy \citep{deason20}.  There currently exist thousands of halo stars with 6D phase-space measurements, thanks to the exquisite  \textit{Gaia} astrometry and wide-field spectroscopic surveys such as the Sloan Digital Sky Survey (SDSS) and the Large Sky Area Multi-Object Fibre Spectroscopic Telescope (LAMOST) survey. Moreover, with future \textit{Gaia} data releases and the next generation of wide-field spectroscopic surveys from facilities such as the Dark Energy Spectroscopic Instrument (DESI, \citealt{desi}), the WHT Enhanced Area Velocity Explorer (WEAVE, \citealt{weave}), and the 4-metre Multi-Object Spectroscopic Telescope (4MOST, \citealt{4most}), there will be hundreds of thousands of halo stars with 6D measurements. The magnitude limit of \textit{Gaia} and the complementary spectroscopic surveys will likely limit the samples of halo stars to within $\sim \! 100$ kpc, but this is still an appreciable fraction of the virial radius ($\sim \! 0.5 r_{\rm 200c}$), and will probe relatively unchartered territory beyond 50 kpc.

As we enter a regime of more precise mass measures, and significantly reduced statistical uncertainties, it is vital to be mindful of any systematic influences in our mass estimates. Although many mass-modelling techniques assume dynamical equilibrium, it is well-documented that ``realistic'' stellar haloes can be a mash-up of several coherent streams and substructures \citep[e.g.][]{bullock05,cooper10, monachesi19}. Thus, comparisons with cosmologically motivated models of stellar haloes are crucial \citep[e.g.][]{yencho06, wang15}. However, while cosmological simulations can provide much needed context,  the \textit{unique} assembly history of the Milky Way is most relevant for Galactic mass measurements. For example, the influence of the Sagittarius (Sgr) stream, which contributes a significant fraction to the total stellar halo mass ($\sim \! 10-15\%$, e.g. \citealt{deason19b}), needs to be considered. Furthermore, and perhaps more importantly, it has recently been recognised that the recent infall of the massive Large Magellanic Cloud (LMC) can imprint significant velocity gradients in the Milky Way halo \citep[e.g.][]{gomez15, erkal19, garavito19,cunningham20,Petersen2020}. Indeed, \cite{erkal20} showed that these velocity gradients can bias equilibrium based mass modelling, and is thus an effect we can no longer ignore.

In this work, we compile a sample of distant ($r > 50$ kpc) halo stars from the literature with 6D phase-space measurements, and use a distribution function analysis to measure the total mass within 100 kpc. We pay particular attention to systematic influences, such as the Sgr stream and the LMC, and, where possible, correct for these perturbative effects. In Section \ref{sec:sample} we describe our sample of halo stars with Galactocentric distances between 50 and 100 kpc. We describe our distribution function analysis in Section \ref{sec:dfs}, and discuss systematic influences, such as unrelaxed substructure and the LMC, in Section \ref{sec:sys}. Our main results are given in Section \ref{sec:results}, and we discuss and conclude in Section \ref{sec:conc}.

\section{Stellar Halo Stars beyond 50 kpc}
\label{sec:sample}
 We have compiled a sample of $N=830$ halo stars with Galactocentric distances between 50 and 100 kpc (see also \citealt{erkal21}). These stars have measured radial velocities and distances, and the vast majority ($\sim \! 98\%$) have proper motion measurements from the \textit{Gaia} early data release 3 \citep[GDR3,][]{egdr3}. Many of these stars derive from large spectroscopic surveys, such as the Sloan Digital Sky Survey (SDSS) and the Large Sky Area Multi-Object Fibre Spectroscopic Telescope (LAMOST) survey. In particular, we use the SDSS blue horizontal branch (BHB) and K giant samples \citep{xue11,xue14} and the LAMOST K Giant sample \citep{yang19}. These are complemented by the following samples in the literature: RR Lyrae (RRL) stars selected from the Palomar Transient Factory with Keck-DEIMOS spectroscopy \citep{cohen17}, and BHB and blue straggler stars selected from SDSS or Hyper Suprime-Cam with VLT-FORS2 spectroscopy \citep{deason12b, belokurov19}. 
 
 After cross matching our sample with GDR3, we find that a number of the sample have, given the distance range we are probing, higher tangential velocities than expected. These are mainly misclassified dwarfs in the K giant samples. Thus, we impose cuts on the total velocity and parallax such that the total velocity is less than 500 km s$^{-1}$ within $1-\sigma$ uncertainty, and $\varpi/\sigma_\varpi < 3$ (see also \citealt{erkal21}).  We also remove $N=27$ (out of $N=123$ in our distance range) of the SDSS BHB sample \citep{xue11}, which \cite{lancaster19} identify as blue straggler contaminants based on their colours and Balmer line shape. The distances of these stars are hence overestimated, and they are likely located at much smaller radii ($<< 50$ kpc). Our final sample of $N=665$ bonafide halo stars in the distance range 50-100 kpc comprises of $N=437$ K giant stars, $N=103$ BHB stars, $N=104$ RRL stars and $N=21$ BS stars. As our sample consists of a range of stellar populations, and is compiled from a variety of sources, the typical uncertainties in radial velocity, distance and proper motion can vary considerably. Typically, the K giants have $5$ km s$^{-1}$ radial velocity errors, 10\% distance errors, and $\sim \! 0.1$ mas yr$^{-1}$ proper motion errors. On the other hand, the BHB and RRL stars typically have $10-20$ km s$^{-1}$ radial velocity errors, 5\% distance errors, and $\sim \! 0.3-0.4$ mas yr$^{-1}$ proper motion errors. In Section \ref{sec:dfs} we describe how these uncertainties are taken into account in our analysis.

 In this work, heliocentric velocities are converted into Galactocentric ones assuming a distance to the Galactic of $r_0=8.122$ kpc \citep{gravity18}, and a circular speed at the position of the Sun of 235 km s$^{-1}$ \citep[e.g.][]{reid04, eilers19}.  Finally, we use the solar peculiar motions derived by \cite{schonrich10}: $(U_\odot, V_\odot, W_\odot)=(11.1,12.24, 7.25)$ km s$^{-1}$. Note that small variations ($<< 10$ km s$^{-1}$) in the assumed solar motion make little different to our main results. We also emphasise that the assumed solar motion is only used to convert heliocentric to Galactocentric quantities: the circular velocity at the position of the Sun is not used in our inference of the Galactic potential (see Sec. \ref{sec:dfs}).
 
\subsection{Sagittarius stream stars}
The presence of un-relaxed substructure is an important consideration in any equilibrium-based mass measurement (see  Section \ref{sec:subs}).  Thus, we identify stars in our sample that likely belong to the Sgr stream, and excise these from our sample. Stars that belong to the Sgr stream are identified from their position on the sky, distance and radial velocity. The Sgr coordinate system defined by \cite{belokurov14} is used to isolate stars close to the Sgr plane ($|B| < 20^\circ$). The predicted distances and radial velocities of Sgr stars along the stream are taken from the results of \cite{hernitschek17}, \cite{belokurov14} and \cite{vasiliev20}. We consider stars that lie within $3-\sigma$ of these tracks (and with $|B| < 20^\circ$) to be Sgr members. This selection is shown explicitly in Figure 2 of \cite{erkal21}. Our procedure identifies $N=182$ Sgr stars (out of $N=665$) in the sample between $50 < r/\mathrm{kpc} < 100$, leaving a total of $N=483$ non-Sgr halo stars.

\section{Power-law Distribution Functions}
\label{sec:dfs}

We model the halo stars using spherical, power-law distribution functions (see \citealt{evans97}). These simple, and flexible distribution functions have been used in several previous works\footnote{An excellent, concise description of these models is given in Section 2 of \cite{eadie16}.} to model halo populations \citep[e.g.][]{deason11,deason12, eadie16, eadie17}. Our motivation for using this approach is two-fold: (i) the spherical, power-law approximation for underlying gravitational potential is a reasonable approximation for NFW-like haloes in this distance range (see e.g. \citealt{watkins10}), and (ii) over a limited radial range we can approximate the stellar halo density profile as a single power-law.  We use a power-law profile for the potential, $\Phi =\Phi_0 \left(r/50 \mathrm{kpc}\right)^{-\gamma}$, where $\gamma$ is constant. To test the above assertion (i) we fit a power-law profile to potentials described by NFW dark matter haloes and a baryonic component appropriate for the Milky Way \citep{bovy13}. For haloes that follow the well-known mass-concentration (with 0.11 dex scatter, see \citealt{dutton14}) relation we find that the overall potential between 50-100 kpc is well described as a power-law with exponent $\gamma=0.5 \pm 0.06$. Indeed, we use this expected $\gamma$ distribution as a prior in our analysis (see also \citealt{eadie19}). Finally, we remark that in the distance range we are probing the adiabatic contraction of the dark matter halo is relatively minor (see \citealt{cautun20}), so we do not consider this effect in this work.

The stellar halo density is also defined as a spherical power-law, with $\rho \propto r^{-\alpha}$. Between 50-100 kpc, most recent studies favour a power-law slope of $\alpha=4$ \citep[e.g.][]{xue15, cohen17, deason18, fukushima19}, and this is the fiducial value we adopt in this work. However, we discuss how changes in this assumption effect our results in Section \ref{sec:results}. The velocity distribution of our model is described in terms of the binding energy $E =
\Phi(r)-\frac{1}{2}(v_l^2+v_b^2+v_{\rm los}^2)$ and the total angular momentum $L = \sqrt{L_x^2+L_y^2+L_z^2}$ as
\begin{equation}
F(E,L) \propto L^{-2\beta} f(E)
\label{eq:even}
\end{equation}
where,
\begin{equation}
\label{eq:df}
f(E) = E^{\beta(\gamma-2)/\gamma+\alpha/\gamma-3/2}
\end{equation}
Here, $\beta$ is the velocity anisotropy of the stellar velocity distribution, which we assume is a constant.

\subsection{Bayesian Inference Method}
We aim to constrain the overall potential and stellar halo velocity anisotropy in the radial range $50 < r/\mathrm{kpc} < 100$. Thus, $\Phi_0$, $\gamma$ and $\beta$ are the parameters we wish to measure from our model.  The most dominant source of uncertainty in the kinematics of the halo stars is the proper motions (and hence the $v_l$ and $v_b$ velocity components), and in some cases there are no proper motion measurements available.  Thus, we reduce the 3D velocity distribution of our model to the line-of-sight velocity distribution (LOSVD) using the following equation:

\begin{eqnarray}
\label{eq:losvd}
F(l,b,d,v_{\rm los}) &=& \int\int E(v_l)E(v_b) \\
                 && \times \:  F(l,b,d,v_{l},v_{b},v_{\rm los})
\mathrm{d}v_l \mathrm{d}v_b, \notag
\end{eqnarray}
where $E(v)$ is the error function, which we assume is a Gaussian with $\sigma(v)=4.74047 \, d \, \sigma(\mu)$. Note, for simplicity, we assume that this Gaussian in $v_l, v_b$ space has no covariance. For cases where there are no measured proper motions $E(v_l)=E(v_b)=1$. A Bayesian inference method is used to derive the unknown parameters, $\Phi_0$, $\gamma$ and $\beta$:
\begin{equation}
\label{eq:ml}
L(\Phi_0,\gamma, \beta)=\sum_{i=1}^N \mathrm{log} F_{\rm los}(l_i,b_i,d_i,v_{\mathrm{los}_i}, \Phi_0,\gamma, \beta),
\end{equation}
Here, $F_{\rm los}$ is the LOSVD (see Eqn \ref{eq:losvd}) and $N$ is the total number of stars in our sample. We use a brute-force grid-based approach, with $\beta \in [0,1]$, $\gamma \in [0.1,1]$ and $\Phi_0 \in [1,20] \times 10^4$ km$^2$ s$^{-2}$. The circular velocity and total mass can be inferred from $\gamma$ and $\Phi_0$: 
\begin{equation}
V^2_{\rm circ} (r) = \gamma \Phi_0 (r/50)^{-\gamma} \\
M(< r) = r V^2_{\rm circ} /G 
\end{equation}
 
Note that the above procedure assumes that the distances and line-of-sight velocities are known, with no appreciable uncertainty. In practice, we perform a four-dimensional integral ($\mu_l, \mu_b, D, v_{\rm los}$) and also marginalise over the distance and line-of-sight velocity dimensions, weighted by their (Gaussian) error distributions.

Finally, we derive posterior distributions for each parameter by adopting a Gaussian prior on $\gamma$ with mean 0.5 and dispersion 0.06, which is appropriate for NFW haloes (see also \citealt{eadie19}). Uniform priors are adopted for the $\Phi_0$ and $\beta$ parameters within the ranges given above.

\section{Systematic effects}
\label{sec:sys}
Our model described in Section \ref{sec:dfs} assumes spherical symmetry, power-law distributions for both the underlying potential and tracer density profile, and a tracer population that is in equilibrium in the gravitational potential. These assumptions, particularly spherical symmetry and equilibrium, are commonplace in many mass-modelling techniques. Here, we discuss two systematic effects that can break these assumptions, and hence systematically affect our inferred mass profile. 

\subsection{The Large Magellanic Cloud}
\label{sec:lmc}
\begin{figure}
    \begin{minipage}{\linewidth}
        \centering
        \includegraphics[width=\textwidth,angle=0]{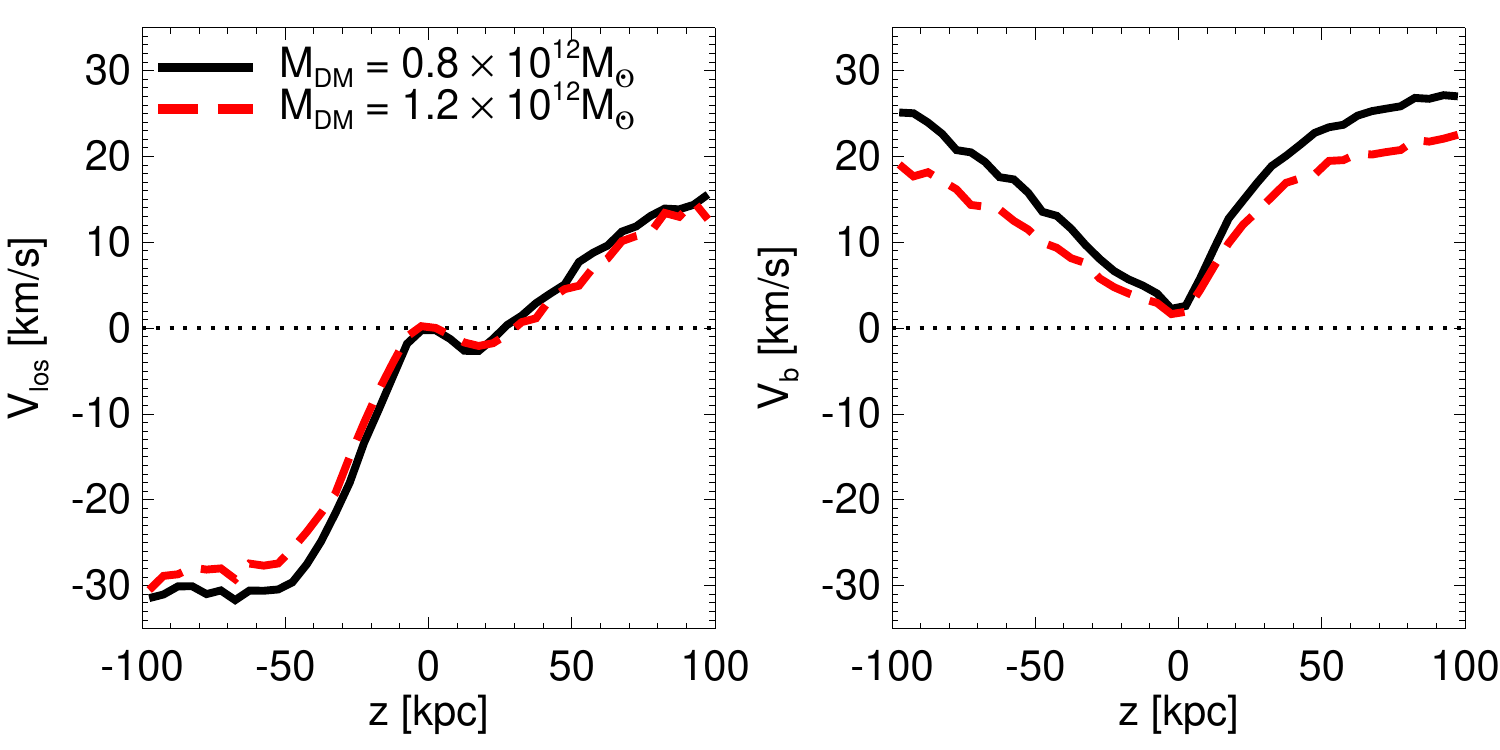}
    \end{minipage}
     \begin{minipage}{\linewidth}
       \centering
           \includegraphics[width=\textwidth,angle=0]{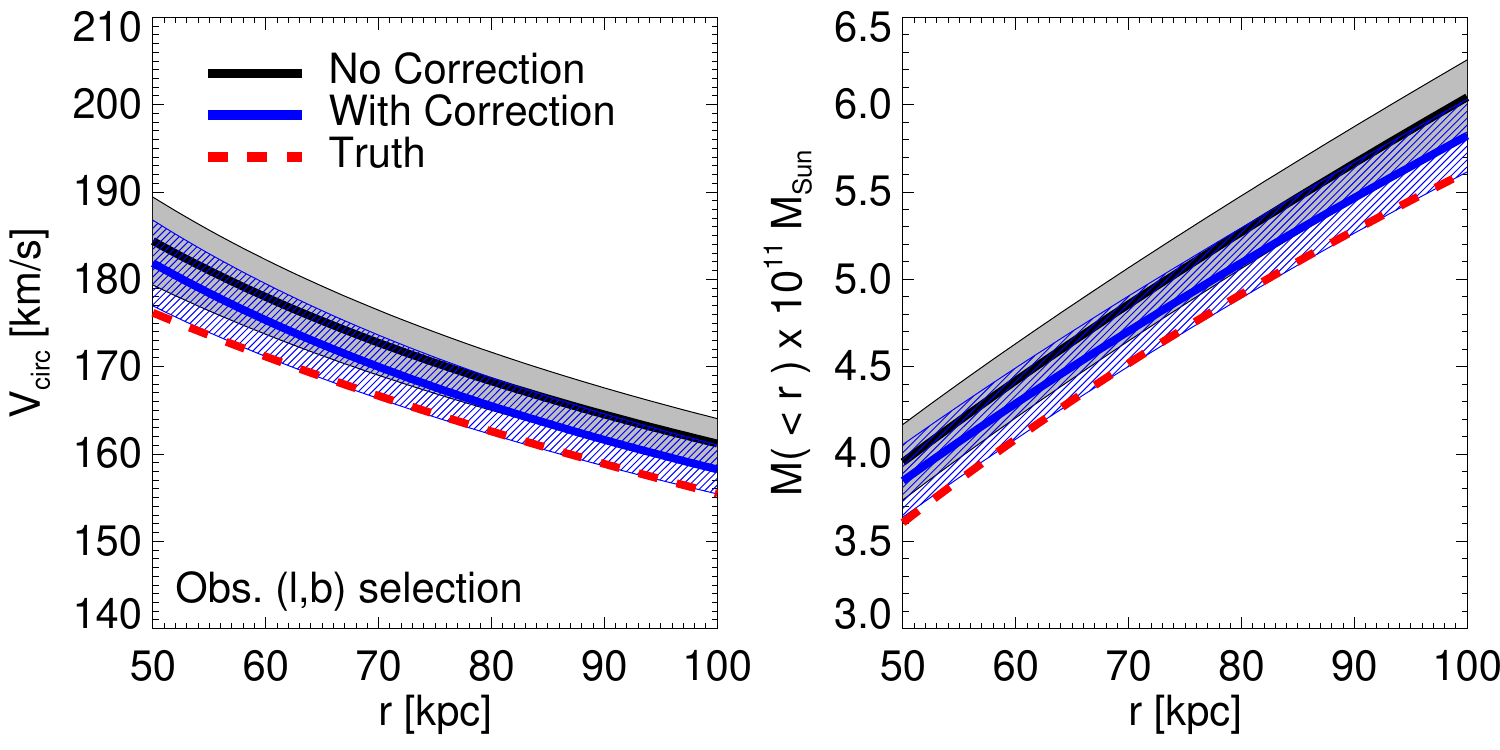}
       \end{minipage}
    \caption[]{\textit{Top panels:} The average $v_{\rm los}$ and $v_b$ velocity components as a function of $z$ (shown in 5 kpc bins) in the \cite{erkal20} stellar halo models with an $1.5 \times 10^{11}$M$_\odot$ LMC. The solid black and dashed red lines are models with $0.8 \times 10^{12}$M$_\odot$ (used in the fiducial model) and $1.2 \times 10^{12}$M$_\odot$ mass dark matter haloes, respectively. Note the $v_l$ component is unaffected by the presence of the LMC. These velocity offsets are used to correct the stellar halo kinematics to approximate equilibrium. \textit{Bottom panels:} The circular velocity and mass profiles between 50-100 kpc derived from a selection of $N =483$ stars from the fiducial \cite{erkal20} model. The stars are selected to have the same Galactic $(l,b)$ distribution as the observed sample. Power-law distribution functions are used to estimate the mass profiles, as described in Section \ref{sec:dfs}. The solid black line shows the resulting profile when no correction to the stellar velocities is applied, and the solid blue line is the result when a correction is applied. The gray shaded and blue line-filled regions indicate the $1-\sigma$ confidence regions. The ``true'' mass profiles are shown with the dashed red line. Note that this ``truth" is the MW mass profile \textit{before} the LMC has been accreted. When a correction is applied our method is able to reproduce the true MW profile within the $1-\sigma$ uncertainty; the mass is typically overestimated when no correction is applied.}
          \label{fig:lmc}
\end{figure}

\begin{figure*}
\centering
        \includegraphics[width=0.95\textwidth,angle=0]{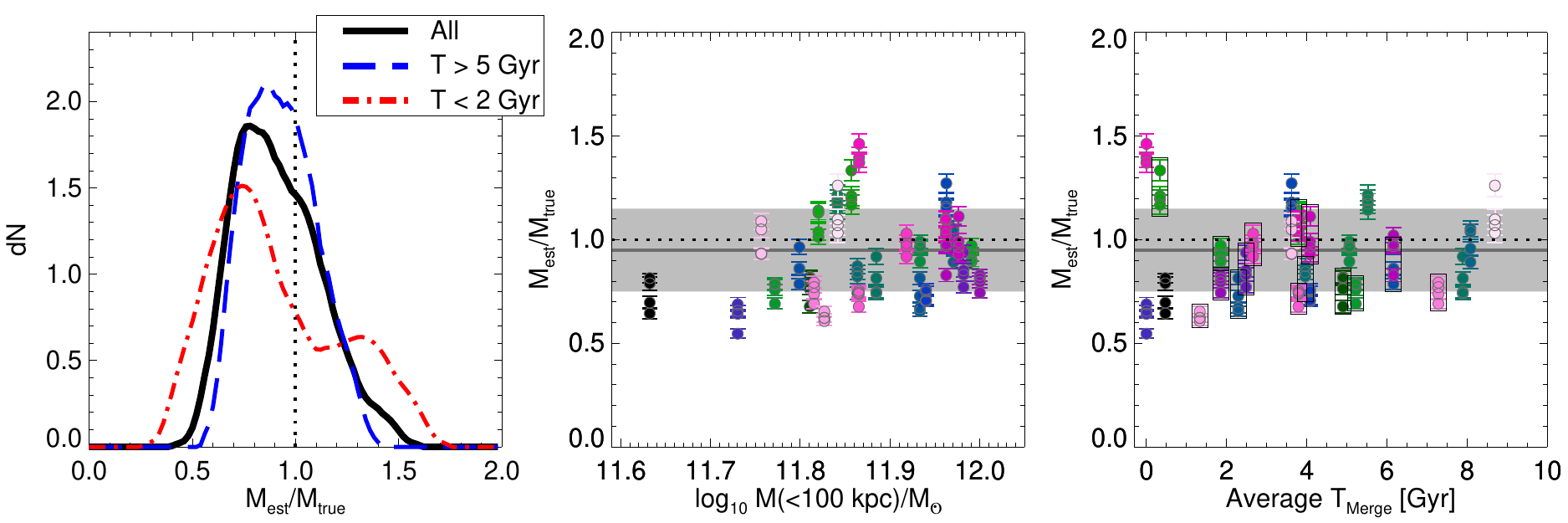}
        \caption[]{The resulting mass estimates (within 100 kpc) when our method is applied to accreted stellar halo stars in the Auriga simulation suite. \textit{Left panel:} The distribution of estimated to true mass ($M_{\rm est} / M_{\rm true} $) for the $N=28$ haloes.  For each halo, we select $N=483$ halo stars with the same Galactic $(l,b)$ distribution as the observed data. This process is repeated four times, each with a different solar position, i.e. $(x,y,z)_{\odot}=(\pm r_0,0,0), (0,\pm r_0,0)$. The $M_{\rm est} / M_{\rm true} $ distribution is smoothed by a Kernel Density Estimation. The dashed blue line and dot-dashed red line show the distributions for haloes with quiescent and active recent accretion, respectively. \textit{Middle panel:} $M_{\rm est} / M_{\rm true} $ against true halo mass within 100 kpc. Each halo is shown with a different colour, where the four values for each halo represent the different solar positions. The grey shaded region indicates a range of $M_{\rm est} / M_{\rm true}$ between 0.75 and 1.15, and the solid line highlights $M_{\rm est} / M_{\rm true} =0.95$ (i.e. $M_{\rm est} / M_{\rm true} = 0.95 \pm 0.20$, see main text). \textit{Right panel:} $M_{\rm est} / M_{\rm true} $ against the average time the progenitor dwarf galaxies that contributed to the halo within 50-100 kpc merged with the host halo. The dispersion of estimated masses is lower at earlier merger times (higher $T_{\rm merge}$), as these deposited stars are more phase-mixed. The black boxes indicate haloes with shell-type structures in the radial range 50-100 kpc. In these cases, the mass tends to be underestimated.}
          \label{fig:auriga}
\end{figure*}

In recent years,  we have realised that the most massive Milky Way satellite, the LMC, has a considerable influence on our Galaxy. This massive satellite ($\sim \! 10^{11}$M$_\odot$) has recently fallen into the Milky Way, and is on a highly eccentric orbit \citep{besla07, kallivayalil13}. The consequences are many-fold, including disturbing the Galactic disc \citep{laporte18}, shifting the barycentre of the Milky Way \citep{gomez15}, and inducing a large scale gravitational wake as the satellite sinks to the centre of the Galaxy due to dynamical friction \citep{garavito19}. These last two considerations are particularly important for halo stars, which, as a result, are imprinted with noticeable velocity gradients \citep{erkal20}. Recently, using the same sample of halo stars as this work, we have detected these predicted velocity gradients in the Galaxy \citep[][see also \citealt{peterson20b}]{erkal21}.

\cite{erkal20} show that the influence of the LMC biases mass estimates of the Milky Way, which assume dynamical equilibrium. In particular, this effect becomes more important at larger distances. We apply our method to the model described in \cite{erkal20} assuming a (rigid) $1.5 \times 10^{11}$M$_\odot$ LMC, which we show in \cite{erkal21} is the most likely LMC mass. This mass is also in close agreement with the LMC mass inferred from perturbations to the Orphan and Sagittarius streams \citep{erkal19,vasiliev20}. The model derives from a suite of simulations of the Milky Way stellar halo in the presence of the LMC, which have been used in previous works \citep[e.g.][]{belokurov19, erkal20}. Note, however, that the deformation of the Milky Way and LMC potential in response to each other, and any resonances in the Milky Way’s dark matter halo, are not included in the models. That said, the simulations do account for the deformation of the stellar halo, and the predictions for stellar halo kinematics using the models in \cite{erkal20} appear to reproduce the salient features in the predictions of the more realistic, fully deforming models in \cite{garavito19}.

We select $N=483$ halo stars from the Erkal et al. model (the same number as our observed sample when Sgr is excluded) between 50-100 kpc, which are chosen to follow the Galactic $(l, b)$ distribution of the observed sample. The slope of the model stellar halo in this radial range is $\alpha=3.4$, and we assume that this slope is known in our analysis. We apply our likelihood method directly to Eqn \ref{eq:even} rather than the LOSVD, and assume all velocity components are known. As we are interested in systematic effects, we do not consider the statistical uncertainties from measurement errors. 

The black lines in the bottom panels of Fig. \ref{fig:lmc} show the resulting circular velocity and mass profiles from this exercise. The ``true" profiles for the model are shown with the dashed red lines. The mass is typically overestimated by $\sim \! 6-7\%$, and the true profile lies outside of the $1-\sigma$ confidence region (shown with the gray shaded region). Note that \cite{erkal20} find a larger bias in this radial range ($\sim \!15\%$). This difference is because the $(l,b)$ distribution of our observed sample is not random on the sky; in particular, a significant fraction of our observed sample ($\sim \!40\%$) lie in the octant on the sky which, according to the Erkal et al. model, is least affected by the presence of the LMC (the octant defined by $(y,z,x)=(-,+,-)$, see Fig. 5 in \citealt{erkal20}).

Although the bias in this case is fairly small, we investigate applying a small correction to minimize this bias. The top panels of Fig. \ref{fig:lmc} show the average $v_{\rm los}$ and $v_b$ velocity components in the \cite{erkal20} model as a function of height above/below the Galactic plane ($z$).  \cite{erkal20} show that the bulk motion of the distant stellar halo is mostly upwards, in the $v_z$ component. Hence, $v_l$ is largely unaffected and the $v_{\rm los}$ and $v_b$ components are shifted. The fiducial model in \cite{erkal20} has an $0.8 \times 10^{12}$M$_\odot$ dark matter halo. We also show the predicted offset for a $1.5 \times$ more massive halo with the red dashed lines. The $v_{\rm los}$ offset is unchanged, and the $v_b$ offset is slightly reduced. We apply a $z$ dependent correction to the $v_{\rm los}$ and $v_b$ velocity components using the trends shown in the top panel of Fig. \ref{fig:lmc}, i.e. $v_{\rm los} = v_{\rm los} - v^{\rm off}_{\rm los}$ and  $v_{\rm b} = v_{\rm b} - v^{\rm off}_{\rm b}$. Here, the relations shown in Fig. \ref{fig:lmc} are given in 5 kpc bins in $z$. We interpolate these values to obtain functions of $v^{\rm off}_{\rm los}(z)$ and $v^{\rm off}_{\rm b}(z)$. Note that we find our results are largely unchanged if we apply an offset appropriate for the more massive halo.  The resulting circular velocity mass profiles when this correction is applied are shown with the blue lines in the bottom panel of the figure. Now, the mass is only overestimated by $\sim \! 3\%$ and the true profile lies within the $1-\sigma$ confidence region.

In Section \ref{sec:results} we apply the same correction to the observational data to approximately account for the reflex motion induced by the LMC. Here, especially when observational errors are taken into account, the correction is relatively small, however, such a procedure will become increasingly more important as we gain larger velocity samples of halo stars over wider regions of the sky. We stress that this correction is only appropriate for the Erkal et al. model of the MW-LMC system, which assumes rigid dark matter potentials. The model dependence of this effect deserves further scrutiny, and, for example, the influence of the LMC on a live MW-LMC system \citep[e.g][]{garavito19} needs to be explored in future work.

Finally, we note that this procedure assumes that the ``true'' Milky Way mass profile does not include the mass of the LMC itself. At $1.5 \times 10^{11}$M$_\odot$ it is clear that this is a significant contribution to the total Milky Way mass. However, the rapid orbit, and likely non-uniform distribution of LMC mass throughout the Galaxy, means that any equilibrium modelling can only hope to correct for the perturbation of the LMC, and recover the mass profile \textit{before} the LMC was accreted. This means that when comparing to the \textit{total} Milky Way mass in cosmological simulations, which includes all the mass within a spherical aperture like the virial radius, the mass of the LMC must be added to the derived equilibrium mass.

\subsection{Unrelaxed Substructure}
\label{sec:subs}

\begin{figure}
  \begin{minipage}{\linewidth}
        \centering
        \includegraphics[width=\textwidth,angle=0]{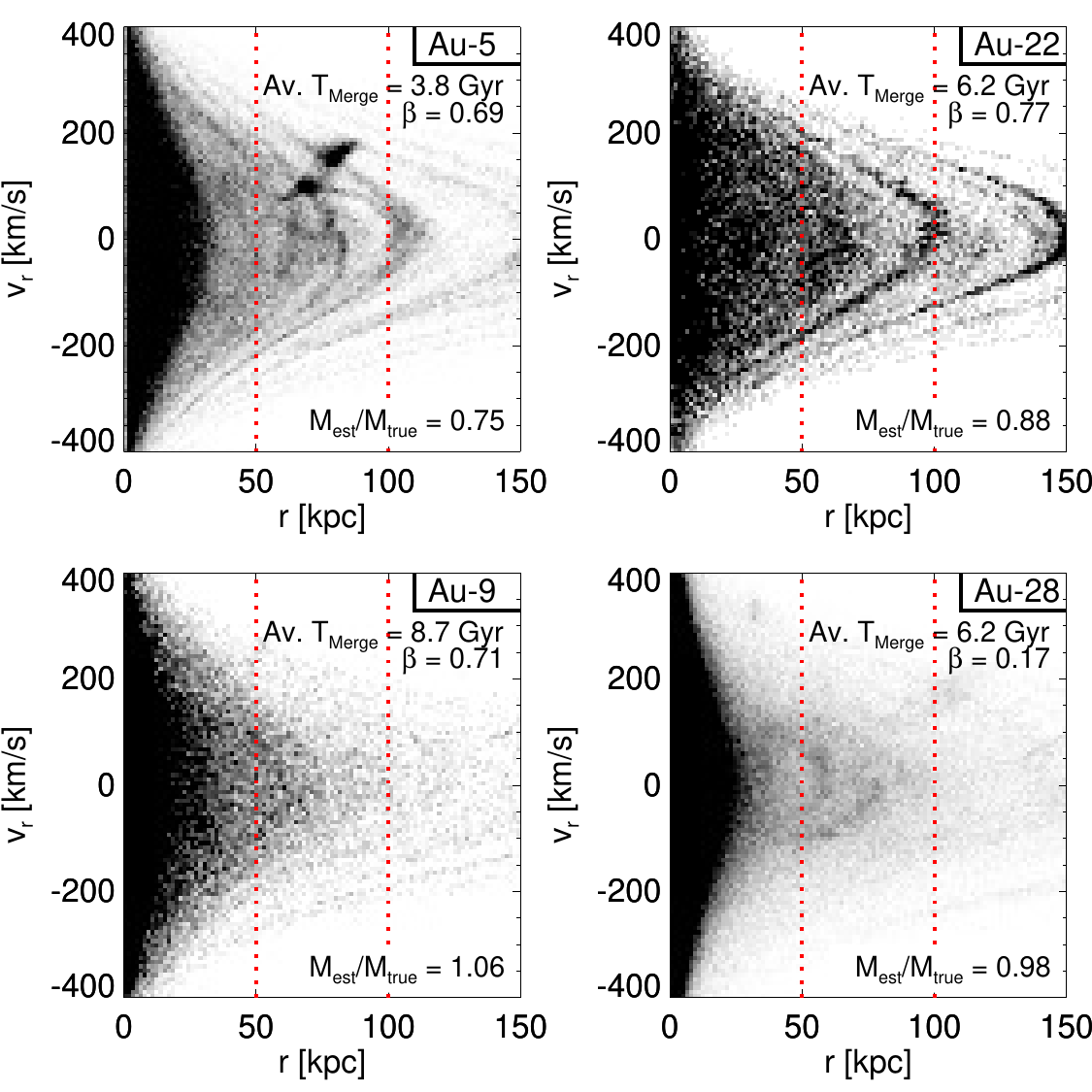}
    \end{minipage}
     \begin{minipage}{\linewidth}
       \centering
           \includegraphics[width=\textwidth,angle=0]{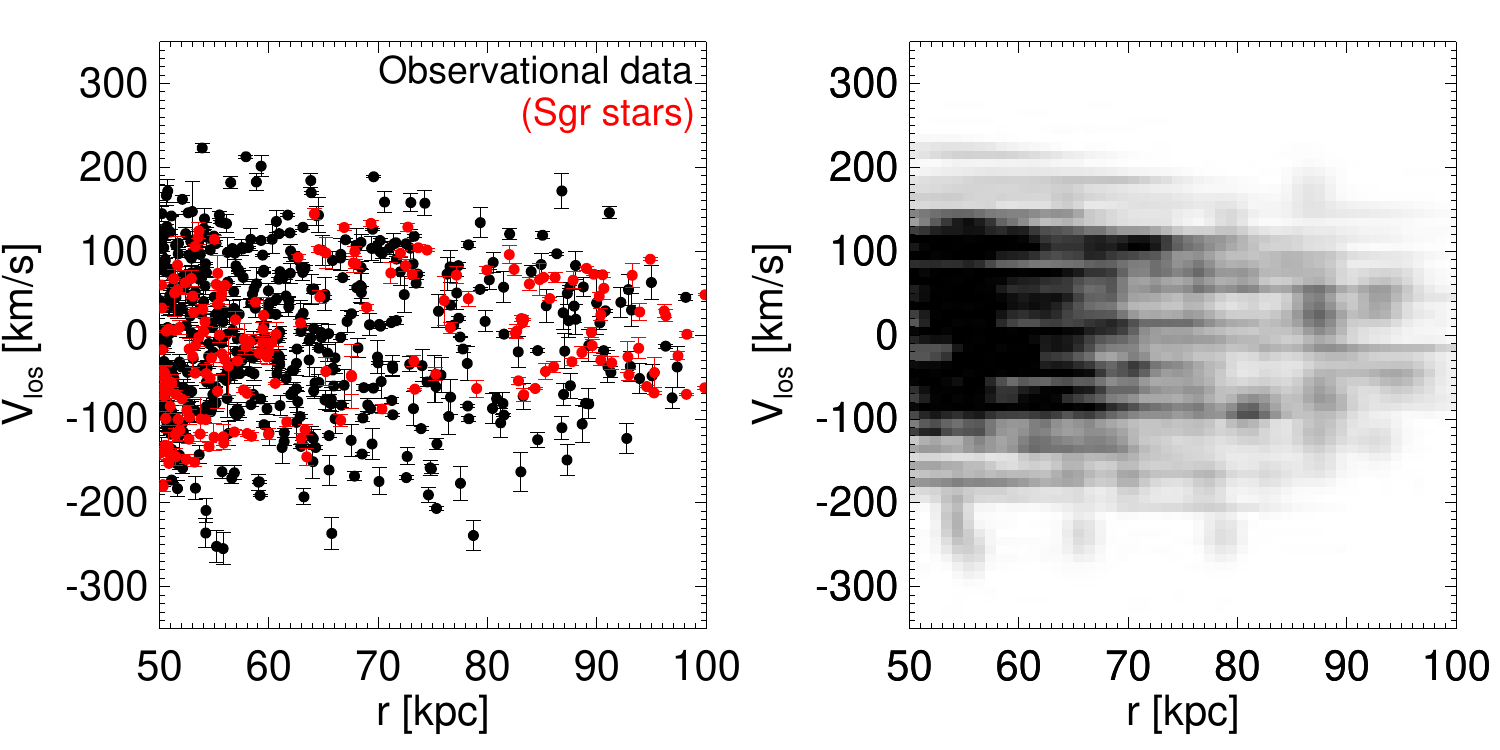}
       \end{minipage}
\centering
        \caption[]{Phase-space diagrams ($v_r$ vs. $r$) for four example Auriga haloes, and the Milky Way data (bottom panels). The top two panels show Auriga haloes with shell-type structures in the radial range $50-100$ kpc. The middle two panels show cases with no obvious shells. Typically, the presence of shells causes the mass estimates to be underestimated. In the bottom two panels we show the $v_{\rm los}$ vs. $r$ diagram for the observational sample in the distance range $50 < r/\mathrm{kpc} < 100$. In the bottom left panel we show each individual star and the associated $v_{\rm los}$ errors. The red points indicate the stars that likely belong to the Sgr stream. The bottom right panel shows a 2D histogram in the $v_{\rm los}$-$r$ space. Here, we have taken into account uncertainties in the distance and velocity of each star. Note that Sgr stars are excluded in the right-hand panel.}
          \label{fig:rvr}
\end{figure}

One only needs to visualise the iconic ``field of streams'' image of the Galactic stellar halo \citep{belokurov06} to see that the stellar halo does not consist of a smooth distribution of stars. Indeed, both observations and simulations have shown that the majority of the (distant) stellar halo is a superposition of the stripped material from destroyed dwarf galaxies; this debris can be lumpy, non-spherical and can have significant velocity gradients. Previous work has investigated how equilibrium modeling holds up under more realistic stellar haloes from simulations \citep{yencho06, eadie18, wang15, han16, sanderson17, wang18, grand19}. Some cases can be wildly inaccurate, but the general consensus is that there is perhaps a systematic floor of at least $\sim 20\%$ accuracy when using halo stars to estimate the total mass. In the past, such a floor has been deemed negligible compared to observational uncertainties. However, we are now in a regime when the data is no longer dominated by statistical uncertainties and/or missing data, and thus this systematic needs to be considered.

Here, we apply our modeling procedure to the halo stars in the Auriga simulations \citep{grand17}.  These are a suite of high resolution, cosmological hydrodynamic simulations of Milky Way-mass galaxies. We only consider the $N=28$ haloes between $M_{200} =1-2 \times 10^{12}$M$_\odot$ that are not currently undergoing a major merger.  As in \cite{fattahi19}, we only consider accreted halo stars, which are defined as those stars that formed in a subhalo other than the main progenitor galaxy. This sample of accreted stars has been used in previous work studying the stellar haloes of the Auriga galaxies \citep[e.g.][]{monachesi16, deason19, monachesi19, fattahi20}. We select $N=483$ halo stars in the radial range 50-100 kpc, and choose these stars to have the same Galactic ($l,b$) distribution as the observational data. For each halo, we create four different samples with different solar positions,  $(x,y,z)_{\odot}=(\pm r_0,0,0), (0,\pm r_0,0)$. The variation between different mass estimates when the solar position is varied is shown in Fig. \ref{fig:auriga}. As in the previous subsection, we only consider systematic effects and do not include observational uncertainties in the analysis. Thus, we assume all of the halo stars have full phase-space information, with no appreciable errors, and we assume that the slope of the tracer density profile ($\alpha$) in this radial range is known.  

In Fig. \ref{fig:auriga} we show the resulting mass estimates (within 100 kpc) when our method is applied to the Auriga simulations. The left panel shows the distribution of estimated to true mass, and the middle panel shows the estimated to true mass against the true mass.  Each halo is shown with different coloured symbols, so the four samples drawn from the same halo (with just the solar position varying) can be identified. The halo-to-halo scatter is significantly higher than the variation from different solar position. The masses are typically underestimated by $\sim \! 10\%$, with a scatter of $\sim \! 25\%$. This bias towards \textit{underestimating} the true mass has been seen in previous work \citep[e.g.][]{yencho06, wang18, grand19}, and the scatter in mass measurements is also in good agreement with previous results. 

The halo-to-halo scatter is much larger than the uncertainties from the likelihood procedure, and thus represents a true systematic effect. To explore this further we calculate the typical time the progenitor dwarf galaxies that contribute to the stars between 50-100 kpc merge with the host galaxy. This is calculated by taking the median accretion time\footnote{Note, here ``accretion'' time refers to when the progenitor dwarf galaxy crosses the virial radius of the host halo.} of all the star particles between 50-100 kpc for each Auriga halo. Here, $T_{\rm merge} =0$ is present day, so more recent accretion events have lower $T_{\rm merge}$ values. The halo-to-halo scatter changes significantly with $T_{\rm merge}$, and the bias is also less pronounced for early time accretion. For example, for debris deposited (on average) over 5 Gyr ago ($T_{\rm merge} > 5$ Gyr), the median of $M_{\rm est}/M_{\rm true}$ is $0.95$ with a dispersion of $0.20$ (see dashed blue line in left hand panel). In contrast, for haloes dominated by material accreted very recently (in the past 2 Gyr, $T_{\rm merge} < 2$ Gyr), the median is $0.82$ with dispersion $0.33$ (see dot-dash red line in left hand panel). The former case, especially when Sgr stars are excluded from the sample, is likely the more appropriate for the Milky Way halo \citep[e.g.][]{ruchti15,deason17, lancaster19,fragkoudi20}.

This exercise shows that the accretion history of the Galaxy is an important consideration for mass-modelling. In particular, if, as is currently believed, the Milky Way has a fairly quiescent recent accretion history, then the mass can be measured to 20\% accuracy with some confidence. However, there still exists a slight (5\%) bias in the mass measurement, which tends to underestimate the halo mass. After visual inspection of the phase-space diagrams of the Auriga haloes (see Fig. \ref{fig:rvr}), we suggest that this bias is due to the presence of shell-like structures in this radial regime \citep[e.g.][]{quinn84}. These shells can form when a massive dwarf galaxy collides with the host, and the stellar debris follows very radial orbits. The enhancement of stars at apocentre thus builds up shell structures. These structures can persist for very long times in the halo \citep[e.g.][]{johnston08}, and are more common at large distances. We are typically probing these structures at apocentre, and their pericentres are found at much lower distances. This leads to an underestimate in the halo mass as we are not fully sampling the phase-space distribution. 

We visually identify that approximately half of the Auriga haloes have shells in the radial regime 50-100 kpc. In the right-hand panel of Fig. \ref{fig:auriga} these cases are indicated with the black boxes. For haloes with $T_{\rm merge} > 2$ Gyr, the median $M_{\rm est}/M_{\rm true}$ is $0.8$ for haloes with shells, and $1.0$ without shells. Four examples of phase-space diagrams are shown in Fig. \ref{fig:rvr}. Here, we show two cases with prominent shells (top panels), and two cases without (middle panels). In the bottom two panels of Fig. \ref{fig:rvr} we show the observational Milky Way data in the $v_{\rm los}$ vs. $r$ plane. In the bottom left panel, each star is shown as a filled circle with its associated velocity error. In the right panel, we show a 2D histogram with pixels of size $1 \mathrm{kpc} \times 10 \mathrm{km s}^{-1}$. Here, we incorporate the observational errors in distance and velocity into the weightings of each pixel to take into account these uncertainties. There is some evidence that there are shell-like features in the data at $r \sim 65-70 $ kpc and $r \sim 80-90$ kpc. However, this is a tentative result because the observational sample does not randomly sample the halo density profile; in fact, the sample is likely biased towards smaller distances, so we must be careful of the interpretation of these features. Nonetheless, there does appear to be a cold features, which warrant further scrutiny with future, more expansive, datasets. Note that the most likely origin of these shells is the highly radial \textit{Gaia}-Enceladus-Sausage merger \citep[e.g.][]{belokurov18, deason18b, helmi18}. Indeed, \cite{fattahi19} showed that the Auriga-5 and Auriga-22 haloes, shown in the top two panels of Fig. \ref{fig:rvr}, have experienced analogous \textit{Gaia}-Enceladus-Sausage merger events.

Interestingly, the bias apparent in the cosmological simulations acts in the opposite sense to the impact of the LMC. Thus, it is crucial that models of the impact of the LMC on a realistic, lumpy halo are explored. For now, in addition to correcting the bias caused by the LMC and excising stars belonging to the Sgr stream, we include a systematic bias of 5\% in our analysis and a systematic uncertainty of 20\% to account for the predictions from the simulations. Here, we assume that the influence of shells could be affecting our mass-estimate, and we adopt the typical bias and uncertainty from the Auriga haloes with relatively quiescent accretion histories. Note, here, we are assuming that the Auriga haloes are a representative sample of MW-mass haloes, and capture the main affects of substructure that are relevant for the MW. In future work, it would be beneficial to compare with much larger samples of high resolution haloes, and test independent simulation suites.

\section{Application to the Milky Way Halo}
\label{sec:results}
\begin{figure}
  \centering
        \includegraphics[width=\linewidth,angle=0]{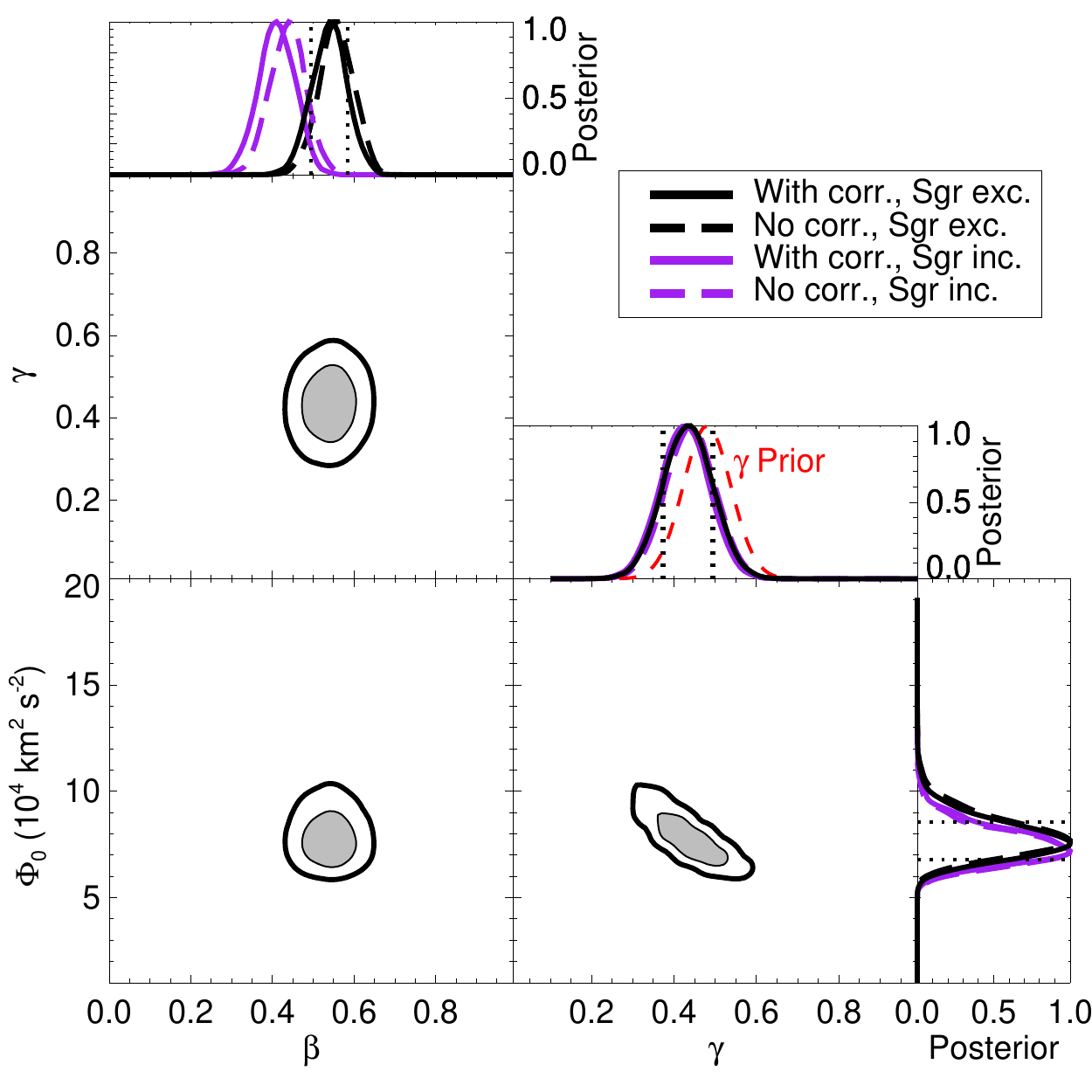}
        \caption[]{The likelihood contours when our method is applied to the observed sample of halo stars between 50-100 kpc. The shaded gray regions and solid black lines indicate the $1-$ and $2-\sigma$ confidence levels, respectively. The inset panels show the 1D posterior distributions for each free parameter in our analysis ($\beta$, $\gamma$ and $\Phi_0$). The prior on the potential slope ($\gamma$) based on NFW halo profiles is shown with the dashed red line. We also show  the resulting posteriors when no velocity offset is used to correct for the influence of the LMC (dashed lines) and when the stars with high probability of belonging to the Sgr stream are included (purple lines).}
          \label{fig:contour}
\end{figure}

We now apply the procedure outlined in Section \ref{sec:dfs} to the Milky Way data. Our sample comprises of $N=665$ halo stars ($N=483$ when Sgr stars are excluded), in the radial range 50-100 kpc. All of the stars have measured radial velocities and distances, and the majority ($98\%$) have proper motion measurements from GDR3. As discussed in Section \ref{sec:dfs}, we apply a Bayesian inference analysis to the power-law distribution function (see eqns. \ref{eq:even} and \ref{eq:df}), which is marginalised over the distance, proper motion, and line-of-sight velocity error distributions.

The resulting likelihood contours and posteriors for the potential parameters ($\Phi_0$, $\gamma$) and velocity anisotropy ($\beta$) are shown in Fig. \ref{fig:contour}. Here, we exclude Sgr stars, and apply an offset to the $v_{\rm los}$ and $v_b$ velocity components to account for the effect of the LMC (see Section \ref{sec:lmc}). We also show the 1D posterior distributions for the cases when: (i) Sgr stars are excluded, but no velocity offset is applied (dashed black), (ii) Sgr stars are included, and a velocity offset is applied (solid purple), and (iii) Sgr stars are included, and no velocity offset is applied (dashed purple). Our results are summarised in Table \ref{tab:like}. We find a radially biased velocity anisotropy, with $\beta = 0.54 \pm 0.05$, in good agreement with \cite{bird19,bird20}, and also in agreement with studies finding that the highly radial \textit{Gaia}-Enceladus-Sausage debris dominates the central regions of the Milky Way \citep[e.g][]{deason18b, lancaster19}. By combining the (highly degenerate) $\Phi_0$ and $\gamma$ parameters, we find the mass within 100 kpc: $M(<100 \mathrm{kpc}) = 5.78 \pm 0.29 \times 10^{11}$M$_\odot$. We note that our prior on $\gamma$ does not significantly affect the total mass measurement within 100 kpc, but is an important influence on the shape of the mass profile that we derive.

\begin{figure}
           \centering
        \includegraphics[width=\linewidth,angle=0]{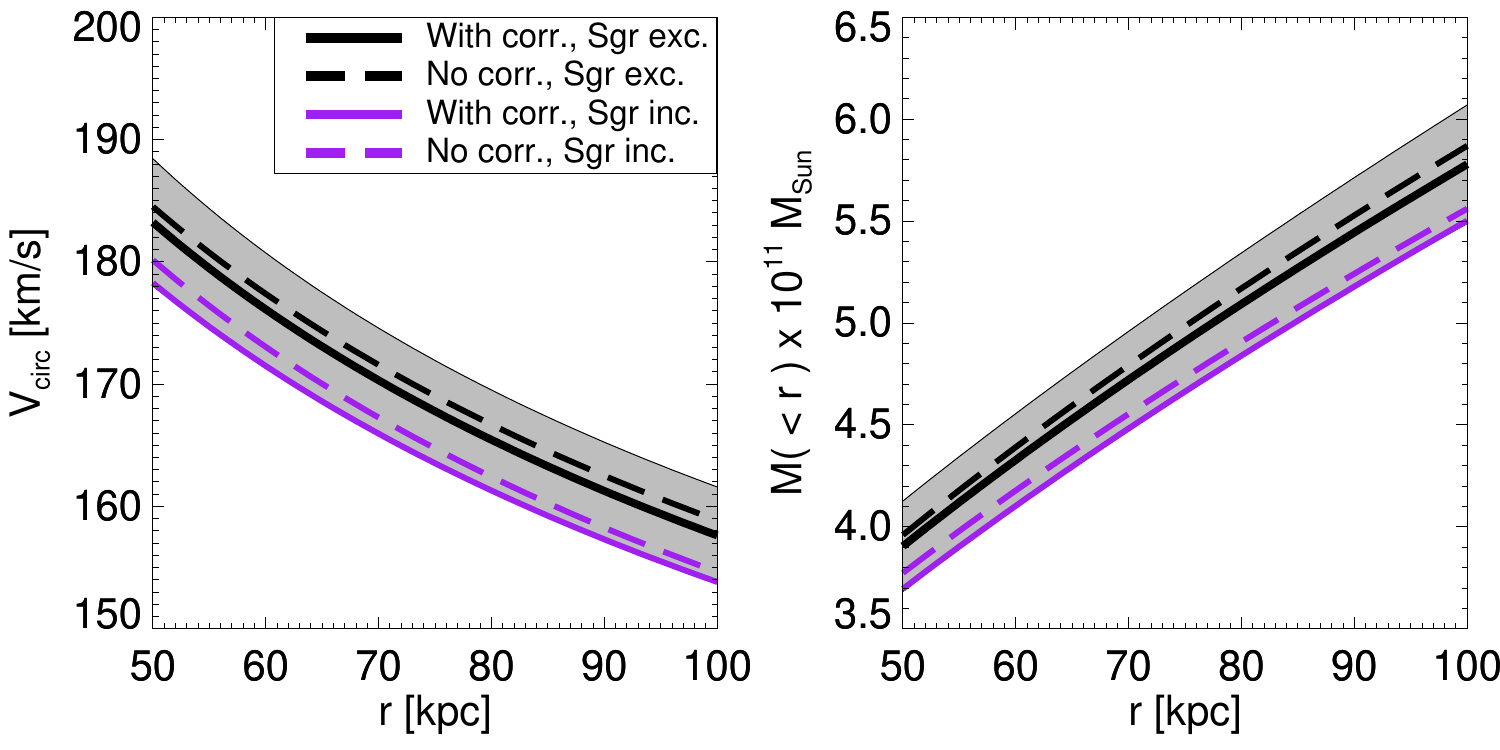}
          \caption[]{The resulting circular velocity (left panel) and mass (right panel) profiles as a function of galactocentric radius. The results when no velocity offset is applied (dashed lines) and when Sgr stars are included (purple lines) are also shown. The shaded regions indicate the $1-\sigma$ uncertainty.}
          \label{fig:mass_profile}
\end{figure}

To further illustrate the influence of excluding Sgr and/or applying a velocity offset we show the resulting circular velocity and mass profiles in Fig. \ref{fig:mass_profile}. The solid black lines and gray shaded regions show the fiducial results (with Sgr exc. and a velocity offset applied) and $1-\sigma$ uncertainty. The dashed lines show the results when an offset is not applied, and the purple lines are when Sgr stars are not excised. Here, we can see that inclusion of Sgr stars biases the mass estimates low, whereas the velocity gradients induced by the LMC bias the mass estimates high. We note, however, that both systematics are relatively small, and within the $1-\sigma$ uncertainties.

In Section \ref{sec:dfs} we discussed how the tracer density slope, $\alpha$, is kept fixed in our analysis. Previous work has shown that the tracer density slope is an important parameter in dynamical mass estimates \citep[e.g.][]{dehnen06, deason12}. Our assumption of $\alpha=4.0$ between $50-100$ kpc is motivated by recent measurements in the literature \citep[e.g.][]{xue15, cohen17, deason18, fukushima19}. However, the stellar halo power-law slope in this radial range is still debated, and the uncertainty is not well quantified. To this end, we provide an approximate fitting formula to adjust our mass measurement based on the input $\alpha$ parameter:
\begin{equation}
M(<100 \,\mathrm{kpc})= M(<100 \, \mathrm{kpc})_{\alpha=4} \times \left(1+0.275\left[\alpha-4.0\right]\right),
\end{equation}
where this equation is valid for $3.0 < \alpha < 5.0$. We emphasize that our fiducial mass estimates in this work use an $\alpha=4.0$ slope. However, the approximate relation given above can, for example, be used to compare with other mass measures that assume a different tracer density profile. In future work, with more expansive halo samples beyond 50 kpc, this parameter will likely be pinned down with much greater confidence. Note that, ideally, the density profile of the halo stars would be a free parameter in our analysis. However, the selection functions of our halo star samples are non-trivial, and ill-defined. While we are confident that these samples do not have kinematic biases, there are undoubtedly magnitude and colour biases that vary across the sky, which hampers our ability to robustly model the halo density profile. However, the more uniform selection functions of upcoming spectroscopic surveys \citep[e.g.][]{desi_sel} will allow us to model the $\alpha$ parameter concurrently with the velocity anisotropy and potential parameters.

\begin{table}
\begin{center}
\renewcommand{\tabcolsep}{0.11cm}
\renewcommand{\arraystretch}{1.5}
\begin{tabular}{| l c  c  c  c |}
 \hline 
Label & $\gamma$ & $\Phi_0$
 & $\beta$ & $M(<100\mathrm{kpc})$\\
& & $[10^4 \mathrm{km}^2\mathrm{s}^{-2}$] & & [$10^{11}$M$_\odot$]\\
 \hline
With corr., Sgr exc. & $0.44^{+0.05}_{-0.07}$ & $7.6^{+1.0}_{-0.8}$ &
$0.54^{+0.05}_{-0.05}$& $5.78 \pm 0.29$ \\
No corr., Sgr exc. & $0.44^{+0.05}_{-0.07}$ & $7.7^{+1.0}_{-0.8}$ &
$0.55^{+0.05}_{-0.04}$& $5.87 \pm 0.29$ \\
With corr., Sgr inc. & $0.43^{+0.06}_{-0.06}$ & $7.2^{+1.0}_{-0.7}$ &
$0.41^{+0.05}_{-0.04}$& $5.50 \pm 0.23$ \\
No corr., Sgr inc. & $0.44^{+0.06}_{-0.06}$ & $7.2^{+0.9}_{-0.7}$ &
$0.44^{+0.04}_{-0.05}$& $5.56 \pm 0.23$ \\
\hline
  \end{tabular}
  \caption{The results of the Bayesian inference analysis. Here, we give the posterior values and associated $1-\sigma$ errors.}
\label{tab:like}
\end{center}
\end{table}

\begin{figure}
  \centering
        \includegraphics[width=\linewidth,angle=0]{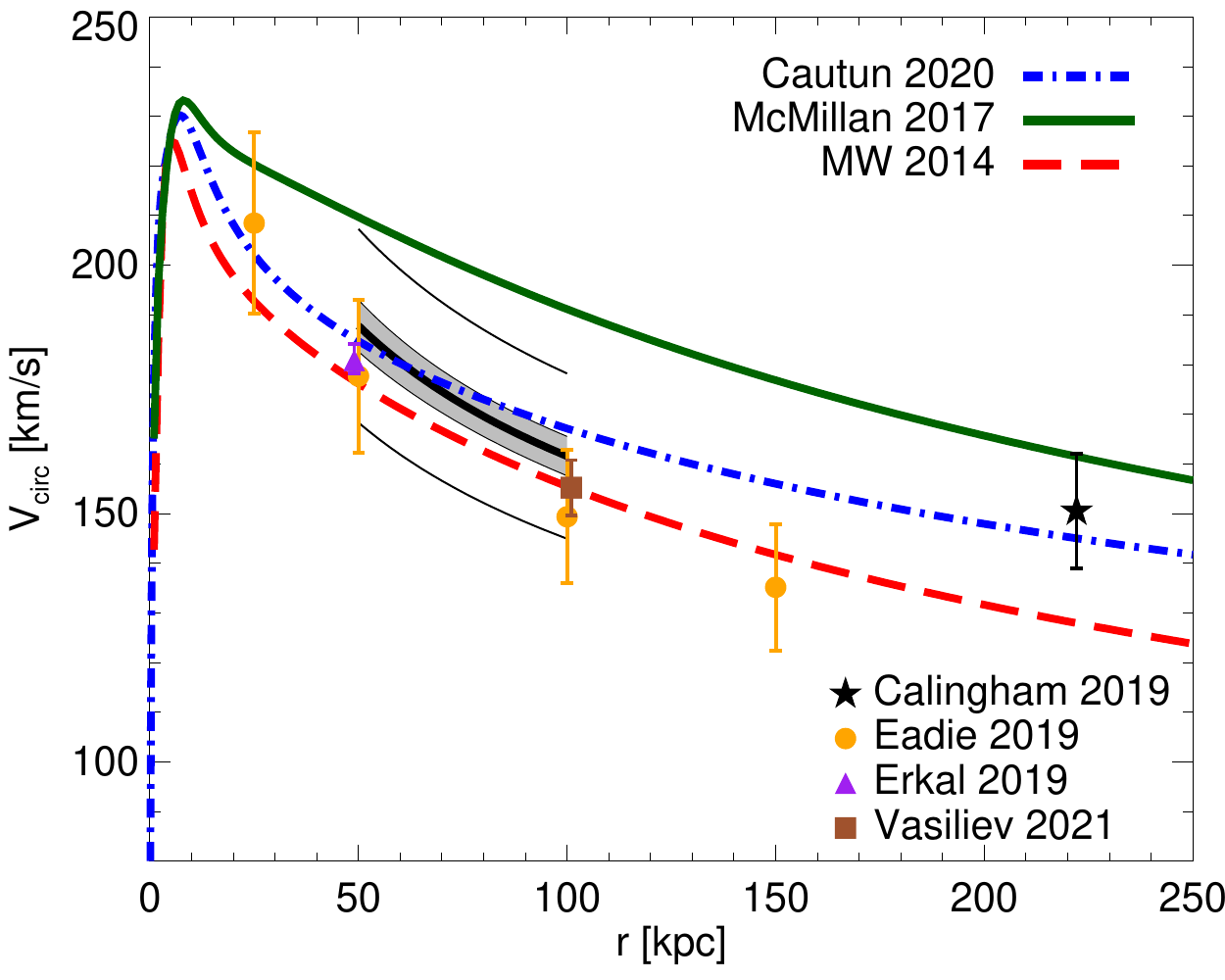}
        \caption[]{Our results in context with other recent mass measurements. Here, we show the circular velocity profile of the Milky Way. Our measurement between 50-100 kpc is shown with the thick black line and the gray shaded region ($1-\sigma$ confidence). The black lines outside of the gray shaded region indicate the overall $1-\sigma$ uncertainty, including an additional 20\% systematic mass uncertainty in the mass measurement (10\% in $V_{\rm circ}$).  The coloured lines show the circular velocity profiles for three commonly used models for the Milky Way: \citep[][dot-dashed blue]{cautun20}; \citep[][solid green]{mcmillan17}; \citep[][dashed red]{bovy15}. We also show recent measurements by \cite{callingham19}, \cite{eadie19}, \cite{erkal19}, and \cite{vasiliev20} with the filled black star, orange circles, purple triangle, and brown square, respectively.}
          \label{fig:vcirc}
\end{figure}

Finally, in Figure \ref{fig:vcirc} we show our derived circular velocity profile between 50-100 kpc in context with other measurements and models in the literature. Here, we have included an additional 20\% systematic uncertainty and a 5\% bias correction, following our tests on the Auriga cosmological simulations: $M(< 100 \mathrm{kpc}) = 6.07 \pm 0.29 \mathrm{(stat.)} \pm 1.21 \mathrm{(sys.)}  \times 10^{11}$M$_\odot$. Assuming the mass-concentration relation for NFW haloes with 0.11 dex scatter \citep{dutton14}, we find that our mass estimate within 100 kpc favours a total Milky Way mass of $M_{\rm 200c} = 1.01 \pm 0.24  \times 10^{12}$M$_\odot$. Note here we assume the baryonic mass given by \cite{bovy13}. As mentioned earlier, the total mass calculated from equilibrium based modeling refers to the Milky Way mass \textit{before} the LMC was accreted. If we include the LMC mass, the total mass (today) is  $M_{\rm 200c} = 1.16 \pm 0.24  \times 10^{12}$M$_\odot$.

Our circular velocity profile shown in Figure \ref{fig:vcirc} is in excellent agreement with the \cite{cautun20} profile. This model uses Milky Way rotation curve data derived from \textit{Gaia} DR2 to fit an adiabatically contracted dark matter halo. Interestingly, \cite{cautun20} find a relatively ``average'' halo concentration for the Milky Way, of $c_{\rm 200c} = 9.4^{+1.9}_{-2.6}$ and a total (pre-LMC infall) mass of $M_{\rm 200c} = 1.08^{+0.20}_{-0.14}  \times 10^{12}$M$_\odot$, in excellent agreement with our total mass estimate. Moreover, our total Milky Way mass also agrees with recent measurement using the more distant satellite galaxies as tracers, \citep[e.g][]{callingham19, li20}. It is reassuring that our results at intermediate radii are in such good agreement with independent measures from both the inner and outer halo.

\section{Conclusions}
\label{sec:conc}
In this work we have applied a distribution function method to a large sample of distant ($r > 50$ kpc) halo stars to estimate the mass of the Milky Way out to 100 kpc. We pay particular attention to the systematic effect of unrelaxed substructure, and the dynamical influence of the LMC. Our main conclusions are summarised as follows:

\begin{itemize}

\item We use a rigid Milky Way-LMC model to constrain the systematic reflex motion effect of the massive LMC on our halo mass estimate. As shown by \cite{erkal20}, the (pre-infall LMC) halo masses are over-predicted due to the velocity gradients caused by the recently infalling LMC. However, we find that a simple velocity offset correction in $v_{\rm los}$ and $v_b$ can minimize the overestimate caused by the reflex motion induced by the LMC, and, assuming a rigid LMC mass of $1.5 \times 10^{11}$ M$_\odot$, we can recover the true mass within $1-\sigma$. 

\item By applying our method to a sample of Milky Way-mass haloes from the Auriga simulation we find that the halo masses are typically underestimated by 10\%. However, this bias is reduced to $\sim \! 5\%$ if we only consider haloes with relatively quiescent recent accretion histories. The residual bias is due to the presence of long-lived shell-like structures in the outer halo. The halo-to-halo scatter is $\sim \!20\%$ for the quiescent haloes, and represents the dominant source of error in the mass estimate of the Milky Way.

\item We apply our distribution function method to $N=483$ halo stars when high probability Sgr stars are excluded. The overall sample has a radial velocity anisotropy, $\beta=0.5$, in good agreement with recent measures in this radial range \citep{bird19,bird20}. Our estimated mass within 100 kpc is $M(< 100 \mathrm{kpc}) = 6.07 \pm 0.29 \mathrm{(stat.)} \pm 1.21 \mathrm{(sys.)}  \times 10^{11}$M$_\odot$. A systematic bias correction ($+5\%$), and additional uncertainty ($20\%$), are included based on our results from the Auriga simulations. The mass estimates are slightly higher when we do not include a velocity offset to correct for the reflex motion induced by the LMC, or slightly lower when Sgr stars are included in our analysis.

\item Our mass estimate within 100 kpc is in good agreement with recent, independent measures in the same radial range \citep[e.g.][]{eadie19, erkal19, vasiliev20}. If we assume the predicted mass-concentration relation for NFW haloes, our measurement favours a total (pre-LMC infall) Milky Way mass of $M_{\rm 200c} = 1.01 \pm 0.24  \times 10^{12}$M$_\odot$, or (post-LMC infall) mass $M_{\rm 200c} = 1.16 \pm 0.24  \times 10^{12}$M$_\odot$ when a rigid $1.5 \times 10^{11}$M$_\odot$ LMC is included. 

\end{itemize}

The sample of halo stars we have used in this work is just the tip of the iceberg in terms of the datasets coming our way in the next few years. We are \textit{already} in a regime where the systematic uncertainties dominate over the statistical uncertainties: this is uncharted territory for mass measurements of the Milky Way, which have, historically, been hampered by incomplete phase-space information. We show in this work that the next phase in mass modelling is to take into account the unique assembly history of the Milky Way in any analysis. In particular, understanding the influence of the LMC, and the dynamics of both relaxed and unrelaxed substructure are crucial. With larger halo samples and more accurate phase-space measurements, these effects will become more and more important. However, there is hope that with a more detailed mapping of the Milky Way's recent accretion history, and more realistic stellar haloes models that include the influence of the live MW-LMC system, we can provide both accurate and precise Milky Way mass measures.

\section*{Acknowledgements}
We thank an anonymous referee for providing useful comments that helped improve the paper. AD thanks Gurtina Besla for valuable comments on the manuscript, and Marius Cautun for providing circular velocity data for Fig. 6. Our gratitude is extended to all of the essential workers that support our livelihood, especially during the Coronavirus pandemic. AD thanks the staff at the Durham University Day Nursery who play a key role in enabling research like this to happen.

AD is supported by a Royal Society University Research Fellowship. AD and AF acknowledge support from the Leverhulme Trust and the Science and Technology Facilities Council (STFC) [grant numbers ST/P000541/1, ST/T000244/1]. FAG acknowledges financial support from CONICYT through the project FONDECYT Regular Nr. 1181264. FAG also acknowledges funding from the Max Planck Society through a Partner Group grant. X. X. Xue, C. Liu, G. Zhao and L. Zhang acknowledge support from the National Natural Science Foundation of China under grants Nos. 11988101, 11873052, 11873057,11890694, 11773033 and National Key R\&D Program of China No. 2019YFA0405500. 

This work used the DiRAC@Durham facility managed by the Institute for Computational Cosmology on behalf of the STFC DiRAC HPC Facility (\url{www.dirac.ac.uk}). The equipment was funded by BEIS capital funding via STFC capital grants ST/K00042X/1, ST/P002293/1, ST/R002371/1 and ST/S002502/1, Durham University and STFC operations grant ST/R000832/1. DiRAC is part of the National e-Infrastructure.

\section*{Data Availability Statement}
The data presented in the figures are available upon request from the corresponding author. 

\bibliographystyle{mnras}
\bibliography{mybib}

\begin{thebibliography}{}
\makeatletter
\relax
\def\mn@urlcharsother{\let\do\@makeother \do\$\do\&\do\#\do\^\do\_\do\%\do\~}
\def\mn@doi{\begingroup\mn@urlcharsother \@ifnextchar [ {\mn@doi@}
  {\mn@doi@[]}}
\def\mn@doi@[#1]#2{\def\@tempa{#1}\ifx\@tempa\@empty \href
  {http://dx.doi.org/#2} {doi:#2}\else \href {http://dx.doi.org/#2} {#1}\fi
  \endgroup}
\def\mn@eprint#1#2{\mn@eprint@#1:#2::\@nil}
\def\mn@eprint@arXiv#1{\href {http://arxiv.org/abs/#1} {{\tt arXiv:#1}}}
\def\mn@eprint@dblp#1{\href {http://dblp.uni-trier.de/rec/bibtex/#1.xml}
  {dblp:#1}}
\def\mn@eprint@#1:#2:#3:#4\@nil{\def\@tempa {#1}\def\@tempb {#2}\def\@tempc
  {#3}\ifx \@tempc \@empty \let \@tempc \@tempb \let \@tempb \@tempa \fi \ifx
  \@tempb \@empty \def\@tempb {arXiv}\fi \@ifundefined
  {mn@eprint@\@tempb}{\@tempb:\@tempc}{\expandafter \expandafter \csname
  mn@eprint@\@tempb\endcsname \expandafter{\@tempc}}}

\bibitem[\protect\citeauthoryear{{Allende Prieto} et~al.,}{{Allende Prieto}
  et~al.}{2020}]{desi_sel}
{Allende Prieto} C.,  et~al., 2020, \mn@doi [Research Notes of the American
  Astronomical Society] {10.3847/2515-5172/abc1dc}, \href
  {https://ui.adsabs.harvard.edu/abs/2020RNAAS...4..188A} {4, 188}

\bibitem[\protect\citeauthoryear{{Belokurov} et~al.,}{{Belokurov}
  et~al.}{2006}]{belokurov06}
{Belokurov} V.,  et~al., 2006, \mn@doi [\apjl] {10.1086/504797}, \href
  {https://ui.adsabs.harvard.edu/abs/2006ApJ...642L.137B} {642, L137}

\bibitem[\protect\citeauthoryear{{Belokurov} et~al.,}{{Belokurov}
  et~al.}{2014}]{belokurov14}
{Belokurov} V.,  et~al., 2014, \mn@doi [\mnras] {10.1093/mnras/stt1862}, \href
  {https://ui.adsabs.harvard.edu/abs/2014MNRAS.437..116B} {437, 116}

\bibitem[\protect\citeauthoryear{{Belokurov}, {Erkal}, {Evans}, {Koposov}  \&
  {Deason}}{{Belokurov} et~al.}{2018}]{belokurov18}
{Belokurov} V.,  {Erkal} D.,  {Evans} N.~W.,  {Koposov} S.~E.,   {Deason}
  A.~J.,  2018, \mn@doi [\mnras] {10.1093/mnras/sty982}, \href
  {https://ui.adsabs.harvard.edu/abs/2018MNRAS.478..611B} {478, 611}

\bibitem[\protect\citeauthoryear{{Belokurov}, {Deason}, {Erkal}, {Koposov},
  {Carballo-Bello}, {Smith}, {Jethwa}  \& {Navarrete}}{{Belokurov}
  et~al.}{2019}]{belokurov19}
{Belokurov} V.,  {Deason} A.~J.,  {Erkal} D.,  {Koposov} S.~E.,
  {Carballo-Bello} J.~A.,  {Smith} M.~C.,  {Jethwa} P.,   {Navarrete} C.,
  2019, \mn@doi [\mnras] {10.1093/mnrasl/slz101}, \href
  {https://ui.adsabs.harvard.edu/abs/2019MNRAS.488L..47B} {488, L47}

\bibitem[\protect\citeauthoryear{{Besla}, {Kallivayalil}, {Hernquist},
  {Robertson}, {Cox}, {van der Marel}  \& {Alcock}}{{Besla}
  et~al.}{2007}]{besla07}
{Besla} G.,  {Kallivayalil} N.,  {Hernquist} L.,  {Robertson} B.,  {Cox} T.~J.,
   {van der Marel} R.~P.,   {Alcock} C.,  2007, \mn@doi [\apj]
  {10.1086/521385}, \href
  {https://ui.adsabs.harvard.edu/abs/2007ApJ...668..949B} {668, 949}

\bibitem[\protect\citeauthoryear{{Bird}, {Xue}, {Liu}, {Shen}, {Flynn}  \&
  {Yang}}{{Bird} et~al.}{2019}]{bird19}
{Bird} S.~A.,  {Xue} X.-X.,  {Liu} C.,  {Shen} J.,  {Flynn} C.,   {Yang} C.,
  2019, \mn@doi [\aj] {10.3847/1538-3881/aafd2e}, \href
  {https://ui.adsabs.harvard.edu/abs/2019AJ....157..104B} {157, 104}

\bibitem[\protect\citeauthoryear{{Bird}, {Xue}, {Liu}, {Shen}, {Flynn}  \&
  {Yang}}{{Bird} et~al.}{2020}]{bird20}
{Bird} S.~A.,  {Xue} X.-X.,  {Liu} C.,  {Shen} J.,  {Flynn} C.,   {Yang} C.,
  2020, arXiv e-prints, \href
  {https://ui.adsabs.harvard.edu/abs/2020arXiv200505980B} {p. arXiv:2005.05980}

\bibitem[\protect\citeauthoryear{{Bland-Hawthorn} \&
  {Gerhard}}{{Bland-Hawthorn} \& {Gerhard}}{2016}]{bland-hawthorn16}
{Bland-Hawthorn} J.,  {Gerhard} O.,  2016, \mn@doi [\araa]
  {10.1146/annurev-astro-081915-023441}, \href
  {https://ui.adsabs.harvard.edu/abs/2016ARA&A..54..529B} {54, 529}

\bibitem[\protect\citeauthoryear{{Bovy}}{{Bovy}}{2015}]{bovy15}
{Bovy} J.,  2015, \mn@doi [\apjs] {10.1088/0067-0049/216/2/29}, \href
  {https://ui.adsabs.harvard.edu/abs/2015ApJS..216...29B} {216, 29}

\bibitem[\protect\citeauthoryear{{Bovy} \& {Rix}}{{Bovy} \&
  {Rix}}{2013}]{bovy13}
{Bovy} J.,  {Rix} H.-W.,  2013, \mn@doi [\apj] {10.1088/0004-637X/779/2/115},
  \href {https://ui.adsabs.harvard.edu/abs/2013ApJ...779..115B} {779, 115}

\bibitem[\protect\citeauthoryear{{Boylan-Kolchin}, {Bullock}  \&
  {Kaplinghat}}{{Boylan-Kolchin} et~al.}{2011}]{boylan-kolchin11}
{Boylan-Kolchin} M.,  {Bullock} J.~S.,   {Kaplinghat} M.,  2011, \mn@doi
  [\mnras] {10.1111/j.1745-3933.2011.01074.x}, \href
  {https://ui.adsabs.harvard.edu/abs/2011MNRAS.415L..40B} {415, L40}

\bibitem[\protect\citeauthoryear{{Bullock} \& {Johnston}}{{Bullock} \&
  {Johnston}}{2005}]{bullock05}
{Bullock} J.~S.,  {Johnston} K.~V.,  2005, \mn@doi [\apj] {10.1086/497422},
  \href {https://ui.adsabs.harvard.edu/abs/2005ApJ...635..931B} {635, 931}

\bibitem[\protect\citeauthoryear{{Callingham} et~al.,}{{Callingham}
  et~al.}{2019}]{callingham19}
{Callingham} T.~M.,  et~al., 2019, \mn@doi [\mnras] {10.1093/mnras/stz365},
  \href {https://ui.adsabs.harvard.edu/abs/2019MNRAS.484.5453C} {484, 5453}

\bibitem[\protect\citeauthoryear{{Cautun} et~al.,}{{Cautun}
  et~al.}{2020}]{cautun20}
{Cautun} M.,  et~al., 2020, \mn@doi [\mnras] {10.1093/mnras/staa1017}, \href
  {https://ui.adsabs.harvard.edu/abs/2020MNRAS.494.4291C} {494, 4291}

\bibitem[\protect\citeauthoryear{{Cohen}, {Sesar}, {Bahnolzer}, {He},
  {Kulkarni}, {Prince}, {Bellm}  \& {Laher}}{{Cohen} et~al.}{2017}]{cohen17}
{Cohen} J.~G.,  {Sesar} B.,  {Bahnolzer} S.,  {He} K.,  {Kulkarni} S.~R.,
  {Prince} T.~A.,  {Bellm} E.,   {Laher} R.~R.,  2017, \mn@doi [\apj]
  {10.3847/1538-4357/aa9120}, \href
  {https://ui.adsabs.harvard.edu/abs/2017ApJ...849..150C} {849, 150}

\bibitem[\protect\citeauthoryear{{Cooper} et~al.,}{{Cooper}
  et~al.}{2010}]{cooper10}
{Cooper} A.~P.,  et~al., 2010, \mn@doi [\mnras]
  {10.1111/j.1365-2966.2010.16740.x}, \href
  {https://ui.adsabs.harvard.edu/abs/2010MNRAS.406..744C} {406, 744}

\bibitem[\protect\citeauthoryear{{Cunningham} et~al.,}{{Cunningham}
  et~al.}{2020}]{cunningham20}
{Cunningham} E.~C.,  et~al., 2020, \mn@doi [\apj] {10.3847/1538-4357/ab9b88},
  \href {https://ui.adsabs.harvard.edu/abs/2020ApJ...898....4C} {898, 4}

\bibitem[\protect\citeauthoryear{{DESI Collaboration} et~al.,}{{DESI
  Collaboration} et~al.}{2016}]{desi}
{DESI Collaboration} et~al., 2016, arXiv e-prints, \href
  {https://ui.adsabs.harvard.edu/abs/2016arXiv161100036D} {p. arXiv:1611.00036}

\bibitem[\protect\citeauthoryear{{Dalton} et~al.,}{{Dalton}
  et~al.}{2014}]{weave}
{Dalton} G.,  et~al., 2014, in \procspie. p. 91470L (\mn@eprint {arXiv}
  {1412.0843}), \mn@doi{10.1117/12.2055132}

\bibitem[\protect\citeauthoryear{{Deason}, {Belokurov}  \& {Evans}}{{Deason}
  et~al.}{2011}]{deason11}
{Deason} A.~J.,  {Belokurov} V.,   {Evans} N.~W.,  2011, \mn@doi [\mnras]
  {10.1111/j.1365-2966.2010.17785.x}, \href
  {https://ui.adsabs.harvard.edu/abs/2011MNRAS.411.1480D} {411, 1480}

\bibitem[\protect\citeauthoryear{{Deason}, {Belokurov}, {Evans}  \&
  {An}}{{Deason} et~al.}{2012a}]{deason12}
{Deason} A.~J.,  {Belokurov} V.,  {Evans} N.~W.,   {An} J.,  2012a, \mn@doi
  [\mnras] {10.1111/j.1745-3933.2012.01283.x}, \href
  {https://ui.adsabs.harvard.edu/abs/2012MNRAS.424L..44D} {424, L44}

\bibitem[\protect\citeauthoryear{{Deason} et~al.,}{{Deason}
  et~al.}{2012b}]{deason12b}
{Deason} A.~J.,  et~al., 2012b, \mn@doi [\mnras]
  {10.1111/j.1365-2966.2012.21639.x}, \href
  {https://ui.adsabs.harvard.edu/abs/2012MNRAS.425.2840D} {425, 2840}

\bibitem[\protect\citeauthoryear{{Deason}, {Belokurov}, {Koposov}, {G{\'o}mez},
  {Grand}, {Marinacci}  \& {Pakmor}}{{Deason} et~al.}{2017}]{deason17}
{Deason} A.~J.,  {Belokurov} V.,  {Koposov} S.~E.,  {G{\'o}mez} F.~A.,  {Grand}
  R.~J.,  {Marinacci} F.,   {Pakmor} R.,  2017, \mn@doi [\mnras]
  {10.1093/mnras/stx1301}, \href
  {https://ui.adsabs.harvard.edu/abs/2017MNRAS.470.1259D} {470, 1259}

\bibitem[\protect\citeauthoryear{{Deason}, {Belokurov}  \& {Koposov}}{{Deason}
  et~al.}{2018a}]{deason18}
{Deason} A.~J.,  {Belokurov} V.,   {Koposov} S.~E.,  2018a, \mn@doi [\apj]
  {10.3847/1538-4357/aa9d19}, \href
  {https://ui.adsabs.harvard.edu/abs/2018ApJ...852..118D} {852, 118}

\bibitem[\protect\citeauthoryear{{Deason}, {Belokurov}, {Koposov}  \&
  {Lancaster}}{{Deason} et~al.}{2018b}]{deason18b}
{Deason} A.~J.,  {Belokurov} V.,  {Koposov} S.~E.,   {Lancaster} L.,  2018b,
  \mn@doi [\apjl] {10.3847/2041-8213/aad0ee}, \href
  {https://ui.adsabs.harvard.edu/abs/2018ApJ...862L...1D} {862, L1}

\bibitem[\protect\citeauthoryear{{Deason}, {Fattahi}, {Belokurov}, {Evans},
  {Grand}, {Marinacci}  \& {Pakmor}}{{Deason} et~al.}{2019a}]{deason19}
{Deason} A.~J.,  {Fattahi} A.,  {Belokurov} V.,  {Evans} N.~W.,  {Grand} R.
  J.~J.,  {Marinacci} F.,   {Pakmor} R.,  2019a, \mn@doi [\mnras]
  {10.1093/mnras/stz623}, \href
  {https://ui.adsabs.harvard.edu/abs/2019MNRAS.485.3514D} {485, 3514}

\bibitem[\protect\citeauthoryear{{Deason}, {Belokurov}  \& {Sanders}}{{Deason}
  et~al.}{2019b}]{deason19b}
{Deason} A.~J.,  {Belokurov} V.,   {Sanders} J.~L.,  2019b, \mn@doi [\mnras]
  {10.1093/mnras/stz2793}, \href
  {https://ui.adsabs.harvard.edu/abs/2019MNRAS.490.3426D} {490, 3426}

\bibitem[\protect\citeauthoryear{{Deason}, {Fattahi}, {Frenk}, {Grand },
  {Oman}, {Garrison-Kimmel}, {Simpson}  \& {Navarro}}{{Deason}
  et~al.}{2020}]{deason20}
{Deason} A.~J.,  {Fattahi} A.,  {Frenk} C.~S.,  {Grand } R. J.~J.,  {Oman}
  K.~A.,  {Garrison-Kimmel} S.,  {Simpson} C.~M.,   {Navarro} J.~F.,  2020,
  \mn@doi [\mnras] {10.1093/mnras/staa1711}, \href
  {https://ui.adsabs.harvard.edu/abs/2020MNRAS.496.3929D} {496, 3929}

\bibitem[\protect\citeauthoryear{{Dehnen}, {McLaughlin}  \&
  {Sachania}}{{Dehnen} et~al.}{2006}]{dehnen06}
{Dehnen} W.,  {McLaughlin} D.~E.,   {Sachania} J.,  2006, \mn@doi [\mnras]
  {10.1111/j.1365-2966.2006.10404.x}, \href
  {https://ui.adsabs.harvard.edu/abs/2006MNRAS.369.1688D} {369, 1688}

\bibitem[\protect\citeauthoryear{{Dutton} \& {Macci{\`o}}}{{Dutton} \&
  {Macci{\`o}}}{2014}]{dutton14}
{Dutton} A.~A.,  {Macci{\`o}} A.~V.,  2014, \mn@doi [\mnras]
  {10.1093/mnras/stu742}, \href
  {https://ui.adsabs.harvard.edu/abs/2014MNRAS.441.3359D} {441, 3359}

\bibitem[\protect\citeauthoryear{{Eadie} \& {Harris}}{{Eadie} \&
  {Harris}}{2016}]{eadie16}
{Eadie} G.~M.,  {Harris} W.~E.,  2016, \mn@doi [\apj]
  {10.3847/0004-637X/829/2/108}, \href
  {https://ui.adsabs.harvard.edu/abs/2016ApJ...829..108E} {829, 108}

\bibitem[\protect\citeauthoryear{{Eadie} \& {Juri{\'c}}}{{Eadie} \&
  {Juri{\'c}}}{2019}]{eadie19}
{Eadie} G.,  {Juri{\'c}} M.,  2019, \mn@doi [\apj] {10.3847/1538-4357/ab0f97},
  \href {https://ui.adsabs.harvard.edu/abs/2019ApJ...875..159E} {875, 159}

\bibitem[\protect\citeauthoryear{{Eadie}, {Springford}  \& {Harris}}{{Eadie}
  et~al.}{2017}]{eadie17}
{Eadie} G.~M.,  {Springford} A.,   {Harris} W.~E.,  2017, \mn@doi [\apj]
  {10.3847/1538-4357/835/2/167}, \href
  {https://ui.adsabs.harvard.edu/abs/2017ApJ...835..167E} {835, 167}

\bibitem[\protect\citeauthoryear{{Eadie}, {Keller}  \& {Harris}}{{Eadie}
  et~al.}{2018}]{eadie18}
{Eadie} G.,  {Keller} B.,   {Harris} W.~E.,  2018, \mn@doi [\apj]
  {10.3847/1538-4357/aadb95}, \href
  {https://ui.adsabs.harvard.edu/abs/2018ApJ...865...72E} {865, 72}

\bibitem[\protect\citeauthoryear{{Eilers}, {Hogg}, {Rix}  \& {Ness}}{{Eilers}
  et~al.}{2019}]{eilers19}
{Eilers} A.-C.,  {Hogg} D.~W.,  {Rix} H.-W.,   {Ness} M.~K.,  2019, \mn@doi
  [\apj] {10.3847/1538-4357/aaf648}, \href
  {https://ui.adsabs.harvard.edu/abs/2019ApJ...871..120E} {871, 120}

\bibitem[\protect\citeauthoryear{{Erkal} et~al.,}{{Erkal}
  et~al.}{2019}]{erkal19}
{Erkal} D.,  et~al., 2019, \mn@doi [\mnras] {10.1093/mnras/stz1371}, \href
  {https://ui.adsabs.harvard.edu/abs/2019MNRAS.487.2685E} {487, 2685}

\bibitem[\protect\citeauthoryear{{Erkal}, {Belokurov}  \& {Parkin}}{{Erkal}
  et~al.}{2020}]{erkal20}
{Erkal} D.,  {Belokurov} V.~A.,   {Parkin} D.~L.,  2020, \mn@doi [\mnras]
  {10.1093/mnras/staa2840}, \href
  {https://ui.adsabs.harvard.edu/abs/2020MNRAS.498.5574E} {498, 5574}

\bibitem[\protect\citeauthoryear{{Erkal} et~al.,}{{Erkal}
  et~al.}{2021}]{erkal21}
{Erkal} D.,  et~al., 2021, arXiv e-prints, \href
  {https://ui.adsabs.harvard.edu/abs/2020arXiv201013789E} {p. arXiv:2010.13789}

\bibitem[\protect\citeauthoryear{{Evans}, {Hafner}  \& {de Zeeuw}}{{Evans}
  et~al.}{1997}]{evans97}
{Evans} N.~W.,  {Hafner} R.~M.,   {de Zeeuw} P.~T.,  1997, \mn@doi [\mnras]
  {10.1093/mnras/286.2.315}, \href
  {https://ui.adsabs.harvard.edu/abs/1997MNRAS.286..315E} {286, 315}

\bibitem[\protect\citeauthoryear{{Fattahi} et~al.,}{{Fattahi}
  et~al.}{2019}]{fattahi19}
{Fattahi} A.,  et~al., 2019, \mn@doi [\mnras] {10.1093/mnras/stz159}, \href
  {https://ui.adsabs.harvard.edu/abs/2019MNRAS.484.4471F} {484, 4471}

\bibitem[\protect\citeauthoryear{{Fattahi} et~al.,}{{Fattahi}
  et~al.}{2020}]{fattahi20}
{Fattahi} A.,  et~al., 2020, \mn@doi [\mnras] {10.1093/mnras/staa2221}, \href
  {https://ui.adsabs.harvard.edu/abs/2020MNRAS.497.4459F} {497, 4459}

\bibitem[\protect\citeauthoryear{{Fragkoudi} et~al.,}{{Fragkoudi}
  et~al.}{2020}]{fragkoudi20}
{Fragkoudi} F.,  et~al., 2020, \mn@doi [\mnras] {10.1093/mnras/staa1104}, \href
  {https://ui.adsabs.harvard.edu/abs/2020MNRAS.494.5936F} {494, 5936}

\bibitem[\protect\citeauthoryear{{Fukushima} et~al.,}{{Fukushima}
  et~al.}{2019}]{fukushima19}
{Fukushima} T.,  et~al., 2019, \mn@doi [\pasj] {10.1093/pasj/psz052}, \href
  {https://ui.adsabs.harvard.edu/abs/2019PASJ...71...72F} {71, 72}

\bibitem[\protect\citeauthoryear{{Gaia Collaboration} et~al.,}{{Gaia
  Collaboration} et~al.}{2018}]{gdr2}
{Gaia Collaboration} et~al., 2018, \mn@doi [\aap]
  {10.1051/0004-6361/201833051}, \href
  {https://ui.adsabs.harvard.edu/abs/2018A&A...616A...1G} {616, A1}

\bibitem[\protect\citeauthoryear{{Gaia Collaboration}, {Brown}, {Vallenari},
  {Prusti}, {de Bruijne}, {Babusiaux}  \& {Biermann}}{{Gaia Collaboration}
  et~al.}{2020}]{egdr3}
{Gaia Collaboration} {Brown} A.~G.~A.,  {Vallenari} A.,  {Prusti} T.,  {de
  Bruijne} J.~H.~J.,  {Babusiaux} C.,   {Biermann} M.,  2020, arXiv e-prints,
  \href {https://ui.adsabs.harvard.edu/abs/2020arXiv201201533G} {p.
  arXiv:2012.01533}

\bibitem[\protect\citeauthoryear{{Garavito-Camargo}, {Besla}, {Laporte},
  {Johnston}, {G{\'o}mez}  \& {Watkins}}{{Garavito-Camargo}
  et~al.}{2019}]{garavito19}
{Garavito-Camargo} N.,  {Besla} G.,  {Laporte} C. F.~P.,  {Johnston} K.~V.,
  {G{\'o}mez} F.~A.,   {Watkins} L.~L.,  2019, \mn@doi [\apj]
  {10.3847/1538-4357/ab32eb}, \href
  {https://ui.adsabs.harvard.edu/abs/2019ApJ...884...51G} {884, 51}

\bibitem[\protect\citeauthoryear{{G{\'o}mez}, {Besla}, {Carpintero},
  {Villalobos}, {O'Shea}  \& {Bell}}{{G{\'o}mez} et~al.}{2015}]{gomez15}
{G{\'o}mez} F.~A.,  {Besla} G.,  {Carpintero} D.~D.,  {Villalobos} {\'A}.,
  {O'Shea} B.~W.,   {Bell} E.~F.,  2015, \mn@doi [\apj]
  {10.1088/0004-637X/802/2/128}, \href
  {https://ui.adsabs.harvard.edu/abs/2015ApJ...802..128G} {802, 128}

\bibitem[\protect\citeauthoryear{{Grand} et~al.,}{{Grand}
  et~al.}{2017}]{grand17}
{Grand} R. J.~J.,  et~al., 2017, \mn@doi [\mnras] {10.1093/mnras/stx071}, \href
  {https://ui.adsabs.harvard.edu/abs/2017MNRAS.467..179G} {467, 179}

\bibitem[\protect\citeauthoryear{{Grand}, {Deason}, {White}, {Simpson},
  {G{\'o}mez}, {Marinacci}  \& {Pakmor}}{{Grand} et~al.}{2019}]{grand19}
{Grand} R. J.~J.,  {Deason} A.~J.,  {White} S. D.~M.,  {Simpson} C.~M.,
  {G{\'o}mez} F.~A.,  {Marinacci} F.,   {Pakmor} R.,  2019, \mn@doi [\mnras]
  {10.1093/mnrasl/slz092}, \href
  {https://ui.adsabs.harvard.edu/abs/2019MNRAS.487L..72G} {487, L72}

\bibitem[\protect\citeauthoryear{{Gravity Collaboration} et~al.,}{{Gravity
  Collaboration} et~al.}{2018}]{gravity18}
{Gravity Collaboration} et~al., 2018, \mn@doi [\aap]
  {10.1051/0004-6361/201833718}, \href
  {https://ui.adsabs.harvard.edu/abs/2018A&A...615L..15G} {615, L15}

\bibitem[\protect\citeauthoryear{{Han}, {Wang}, {Cole}  \& {Frenk}}{{Han}
  et~al.}{2016}]{han16}
{Han} J.,  {Wang} W.,  {Cole} S.,   {Frenk} C.~S.,  2016, \mn@doi [\mnras]
  {10.1093/mnras/stv2522}, \href
  {https://ui.adsabs.harvard.edu/abs/2016MNRAS.456.1017H} {456, 1017}

\bibitem[\protect\citeauthoryear{{Helmi}, {Babusiaux}, {Koppelman}, {Massari},
  {Veljanoski}  \& {Brown}}{{Helmi} et~al.}{2018}]{helmi18}
{Helmi} A.,  {Babusiaux} C.,  {Koppelman} H.~H.,  {Massari} D.,  {Veljanoski}
  J.,   {Brown} A. G.~A.,  2018, \mn@doi [\nat] {10.1038/s41586-018-0625-x},
  \href {https://ui.adsabs.harvard.edu/abs/2018Natur.563...85H} {563, 85}

\bibitem[\protect\citeauthoryear{{Hernitschek} et~al.,}{{Hernitschek}
  et~al.}{2017}]{hernitschek17}
{Hernitschek} N.,  et~al., 2017, \mn@doi [\apj] {10.3847/1538-4357/aa960c},
  \href {https://ui.adsabs.harvard.edu/abs/2017ApJ...850...96H} {850, 96}

\bibitem[\protect\citeauthoryear{{Johnston}, {Bullock}, {Sharma}, {Font},
  {Robertson}  \& {Leitner}}{{Johnston} et~al.}{2008}]{johnston08}
{Johnston} K.~V.,  {Bullock} J.~S.,  {Sharma} S.,  {Font} A.,  {Robertson}
  B.~E.,   {Leitner} S.~N.,  2008, \mn@doi [\apj] {10.1086/592228}, \href
  {https://ui.adsabs.harvard.edu/abs/2008ApJ...689..936J} {689, 936}

\bibitem[\protect\citeauthoryear{{Kallivayalil}, {van der Marel}, {Besla},
  {Anderson}  \& {Alcock}}{{Kallivayalil} et~al.}{2013}]{kallivayalil13}
{Kallivayalil} N.,  {van der Marel} R.~P.,  {Besla} G.,  {Anderson} J.,
  {Alcock} C.,  2013, \mn@doi [\apj] {10.1088/0004-637X/764/2/161}, \href
  {https://ui.adsabs.harvard.edu/abs/2013ApJ...764..161K} {764, 161}

\bibitem[\protect\citeauthoryear{{Kennedy}, {Frenk}, {Cole}  \&
  {Benson}}{{Kennedy} et~al.}{2014}]{kennedy14}
{Kennedy} R.,  {Frenk} C.,  {Cole} S.,   {Benson} A.,  2014, \mn@doi [\mnras]
  {10.1093/mnras/stu719}, \href
  {https://ui.adsabs.harvard.edu/abs/2014MNRAS.442.2487K} {442, 2487}

\bibitem[\protect\citeauthoryear{{Lancaster}, {Belokurov}  \&
  {Evans}}{{Lancaster} et~al.}{2019}]{lancaster19}
{Lancaster} L.,  {Belokurov} V.,   {Evans} N.~W.,  2019, \mn@doi [\mnras]
  {10.1093/mnras/stz124}, \href
  {https://ui.adsabs.harvard.edu/abs/2019MNRAS.484.2556L} {484, 2556}

\bibitem[\protect\citeauthoryear{{Laporte}, {Johnston}, {G{\'o}mez},
  {Garavito-Camargo}  \& {Besla}}{{Laporte} et~al.}{2018}]{laporte18}
{Laporte} C. F.~P.,  {Johnston} K.~V.,  {G{\'o}mez} F.~A.,  {Garavito-Camargo}
  N.,   {Besla} G.,  2018, \mn@doi [\mnras] {10.1093/mnras/sty1574}, \href
  {https://ui.adsabs.harvard.edu/abs/2018MNRAS.481..286L} {481, 286}

\bibitem[\protect\citeauthoryear{{Li}, {Qian}, {Han}, {Li}, {Wang}  \&
  {Jing}}{{Li} et~al.}{2020}]{li20}
{Li} Z.-Z.,  {Qian} Y.-Z.,  {Han} J.,  {Li} T.~S.,  {Wang} W.,   {Jing} Y.~P.,
  2020, \mn@doi [\apj] {10.3847/1538-4357/ab84f0}, \href
  {https://ui.adsabs.harvard.edu/abs/2020ApJ...894...10L} {894, 10}

\bibitem[\protect\citeauthoryear{{Lovell}, {Frenk}, {Eke}, {Jenkins}, {Gao}  \&
  {Theuns}}{{Lovell} et~al.}{2014}]{lovell14}
{Lovell} M.~R.,  {Frenk} C.~S.,  {Eke} V.~R.,  {Jenkins} A.,  {Gao} L.,
  {Theuns} T.,  2014, \mn@doi [\mnras] {10.1093/mnras/stt2431}, \href
  {https://ui.adsabs.harvard.edu/abs/2014MNRAS.439..300L} {439, 300}

\bibitem[\protect\citeauthoryear{{McMillan}}{{McMillan}}{2017}]{mcmillan17}
{McMillan} P.~J.,  2017, \mn@doi [\mnras] {10.1093/mnras/stw2759}, \href
  {https://ui.adsabs.harvard.edu/abs/2017MNRAS.465...76M} {465, 76}

\bibitem[\protect\citeauthoryear{{Metz}, {Kroupa}  \& {Jerjen}}{{Metz}
  et~al.}{2007}]{metz07}
{Metz} M.,  {Kroupa} P.,   {Jerjen} H.,  2007, \mn@doi [\mnras]
  {10.1111/j.1365-2966.2006.11228.x}, \href
  {https://ui.adsabs.harvard.edu/abs/2007MNRAS.374.1125M} {374, 1125}

\bibitem[\protect\citeauthoryear{{Monachesi}, {G{\'o}mez}, {Grand },
  {Kauffmann}, {Marinacci}, {Pakmor}, {Springel}  \& {Frenk}}{{Monachesi}
  et~al.}{2016}]{monachesi16}
{Monachesi} A.,  {G{\'o}mez} F.~A.,  {Grand } R. J.~J.,  {Kauffmann} G.,
  {Marinacci} F.,  {Pakmor} R.,  {Springel} V.,   {Frenk} C.~S.,  2016, \mn@doi
  [\mnras] {10.1093/mnrasl/slw052}, \href
  {https://ui.adsabs.harvard.edu/abs/2016MNRAS.459L..46M} {459, L46}

\bibitem[\protect\citeauthoryear{{Monachesi} et~al.,}{{Monachesi}
  et~al.}{2019}]{monachesi19}
{Monachesi} A.,  et~al., 2019, \mn@doi [\mnras] {10.1093/mnras/stz538}, \href
  {https://ui.adsabs.harvard.edu/abs/2019MNRAS.485.2589M} {485, 2589}

\bibitem[\protect\citeauthoryear{{Moore}, {Ghigna}, {Governato}, {Lake},
  {Quinn}, {Stadel}  \& {Tozzi}}{{Moore} et~al.}{1999}]{moore99}
{Moore} B.,  {Ghigna} S.,  {Governato} F.,  {Lake} G.,  {Quinn} T.,  {Stadel}
  J.,   {Tozzi} P.,  1999, \mn@doi [\apjl] {10.1086/312287}, \href
  {https://ui.adsabs.harvard.edu/abs/1999ApJ...524L..19M} {524, L19}

\bibitem[\protect\citeauthoryear{{Patel}, {Besla}, {Mandel}  \& {Sohn}}{{Patel}
  et~al.}{2018}]{patel18}
{Patel} E.,  {Besla} G.,  {Mandel} K.,   {Sohn} S.~T.,  2018, \mn@doi [\apj]
  {10.3847/1538-4357/aab78f}, \href
  {https://ui.adsabs.harvard.edu/abs/2018ApJ...857...78P} {857, 78}

\bibitem[\protect\citeauthoryear{{Petersen} \& {Pe{\~n}arrubia}}{{Petersen} \&
  {Pe{\~n}arrubia}}{2020a}]{peterson20b}
{Petersen} M.~S.,  {Pe{\~n}arrubia} J.,  2020a, \mn@doi [Nature Astronomy]
  {10.1038/s41550-020-01254-3}, \href
  {https://ui.adsabs.harvard.edu/abs/2020NatAs.tmp..236P} {}

\bibitem[\protect\citeauthoryear{{Petersen} \& {Pe{\~n}arrubia}}{{Petersen} \&
  {Pe{\~n}arrubia}}{2020b}]{Petersen2020}
{Petersen} M.~S.,  {Pe{\~n}arrubia} J.,  2020b, \mn@doi [\mnras]
  {10.1093/mnrasl/slaa029}, \href
  {https://ui.adsabs.harvard.edu/abs/2020MNRAS.494L..11P} {494, L11}

\bibitem[\protect\citeauthoryear{{Quinn}}{{Quinn}}{1984}]{quinn84}
{Quinn} P.~J.,  1984, \mn@doi [\apj] {10.1086/161924}, \href
  {https://ui.adsabs.harvard.edu/abs/1984ApJ...279..596Q} {279, 596}

\bibitem[\protect\citeauthoryear{{Reid} \& {Brunthaler}}{{Reid} \&
  {Brunthaler}}{2004}]{reid04}
{Reid} M.~J.,  {Brunthaler} A.,  2004, \mn@doi [\apj] {10.1086/424960}, \href
  {https://ui.adsabs.harvard.edu/abs/2004ApJ...616..872R} {616, 872}

\bibitem[\protect\citeauthoryear{{Ruchti} et~al.,}{{Ruchti}
  et~al.}{2015}]{ruchti15}
{Ruchti} G.~R.,  et~al., 2015, \mn@doi [\mnras] {10.1093/mnras/stv807}, \href
  {https://ui.adsabs.harvard.edu/abs/2015MNRAS.450.2874R} {450, 2874}

\bibitem[\protect\citeauthoryear{{Sanderson}, {Hartke}  \& {Helmi}}{{Sanderson}
  et~al.}{2017}]{sanderson17}
{Sanderson} R.~E.,  {Hartke} J.,   {Helmi} A.,  2017, \mn@doi [\apj]
  {10.3847/1538-4357/aa5eb4}, \href
  {https://ui.adsabs.harvard.edu/abs/2017ApJ...836..234S} {836, 234}

\bibitem[\protect\citeauthoryear{{Sch{\"o}nrich}, {Binney}  \&
  {Dehnen}}{{Sch{\"o}nrich} et~al.}{2010}]{schonrich10}
{Sch{\"o}nrich} R.,  {Binney} J.,   {Dehnen} W.,  2010, \mn@doi [\mnras]
  {10.1111/j.1365-2966.2010.16253.x}, \href
  {https://ui.adsabs.harvard.edu/abs/2010MNRAS.403.1829S} {403, 1829}

\bibitem[\protect\citeauthoryear{{Vasiliev}}{{Vasiliev}}{2019}]{vasiliev19}
{Vasiliev} E.,  2019, \mn@doi [\mnras] {10.1093/mnras/stz171}, \href
  {https://ui.adsabs.harvard.edu/abs/2019MNRAS.484.2832V} {484, 2832}

\bibitem[\protect\citeauthoryear{{Vasiliev}, {Belokurov}  \&
  {Erkal}}{{Vasiliev} et~al.}{2021}]{vasiliev20}
{Vasiliev} E.,  {Belokurov} V.,   {Erkal} D.,  2021, \mn@doi [\mnras]
  {10.1093/mnras/staa3673}, \href
  {https://ui.adsabs.harvard.edu/abs/2021MNRAS.501.2279V} {501, 2279}

\bibitem[\protect\citeauthoryear{{Wang}, {Frenk}, {Navarro}, {Gao}  \&
  {Sawala}}{{Wang} et~al.}{2012}]{wang12}
{Wang} J.,  {Frenk} C.~S.,  {Navarro} J.~F.,  {Gao} L.,   {Sawala} T.,  2012,
  \mn@doi [\mnras] {10.1111/j.1365-2966.2012.21357.x}, \href
  {https://ui.adsabs.harvard.edu/abs/2012MNRAS.424.2715W} {424, 2715}

\bibitem[\protect\citeauthoryear{{Wang}, {Han}, {Cooper}, {Cole}, {Frenk}  \&
  {Lowing}}{{Wang} et~al.}{2015}]{wang15}
{Wang} W.,  {Han} J.,  {Cooper} A.~P.,  {Cole} S.,  {Frenk} C.,   {Lowing} B.,
  2015, \mn@doi [\mnras] {10.1093/mnras/stv1647}, \href
  {https://ui.adsabs.harvard.edu/abs/2015MNRAS.453..377W} {453, 377}

\bibitem[\protect\citeauthoryear{{Wang}, {Han}, {Cole}, {More}, {Frenk}  \&
  {Schaller}}{{Wang} et~al.}{2018}]{wang18}
{Wang} W.,  {Han} J.,  {Cole} S.,  {More} S.,  {Frenk} C.,   {Schaller} M.,
  2018, \mn@doi [\mnras] {10.1093/mnras/sty706}, \href
  {https://ui.adsabs.harvard.edu/abs/2018MNRAS.476.5669W} {476, 5669}

\bibitem[\protect\citeauthoryear{{Wang}, {Han}, {Cautun}, {Li}  \&
  {Ishigaki}}{{Wang} et~al.}{2020}]{wang19}
{Wang} W.,  {Han} J.,  {Cautun} M.,  {Li} Z.,   {Ishigaki} M.~N.,  2020,
  \mn@doi [Science China Physics, Mechanics, and Astronomy]
  {10.1007/s11433-019-1541-6}, \href
  {https://ui.adsabs.harvard.edu/abs/2020SCPMA..63j9801W} {63, 109801}

\bibitem[\protect\citeauthoryear{{Watkins}, {Evans}  \& {An}}{{Watkins}
  et~al.}{2010}]{watkins10}
{Watkins} L.~L.,  {Evans} N.~W.,   {An} J.~H.,  2010, \mn@doi [\mnras]
  {10.1111/j.1365-2966.2010.16708.x}, \href
  {https://ui.adsabs.harvard.edu/abs/2010MNRAS.406..264W} {406, 264}

\bibitem[\protect\citeauthoryear{{Watkins}, {van der Marel}, {Sohn}  \&
  {Evans}}{{Watkins} et~al.}{2019}]{watkins19}
{Watkins} L.~L.,  {van der Marel} R.~P.,  {Sohn} S.~T.,   {Evans} N.~W.,  2019,
  \mn@doi [\apj] {10.3847/1538-4357/ab089f}, \href
  {https://ui.adsabs.harvard.edu/abs/2019ApJ...873..118W} {873, 118}

\bibitem[\protect\citeauthoryear{{Xue} et~al.,}{{Xue} et~al.}{2011}]{xue11}
{Xue} X.-X.,  et~al., 2011, \mn@doi [\apj] {10.1088/0004-637X/738/1/79}, \href
  {https://ui.adsabs.harvard.edu/abs/2011ApJ...738...79X} {738, 79}

\bibitem[\protect\citeauthoryear{{Xue} et~al.,}{{Xue} et~al.}{2014}]{xue14}
{Xue} X.-X.,  et~al., 2014, \mn@doi [\apj] {10.1088/0004-637X/784/2/170}, \href
  {https://ui.adsabs.harvard.edu/abs/2014ApJ...784..170X} {784, 170}

\bibitem[\protect\citeauthoryear{{Xue}, {Rix}, {Ma}, {Morrison}, {Bovy},
  {Sesar}  \& {Janesh}}{{Xue} et~al.}{2015}]{xue15}
{Xue} X.-X.,  {Rix} H.-W.,  {Ma} Z.,  {Morrison} H.,  {Bovy} J.,  {Sesar} B.,
  {Janesh} W.,  2015, \mn@doi [\apj] {10.1088/0004-637X/809/2/144}, \href
  {https://ui.adsabs.harvard.edu/abs/2015ApJ...809..144X} {809, 144}

\bibitem[\protect\citeauthoryear{{Yang} et~al.,}{{Yang} et~al.}{2019}]{yang19}
{Yang} C.,  et~al., 2019, \mn@doi [\apj] {10.3847/1538-4357/ab2462}, \href
  {https://ui.adsabs.harvard.edu/abs/2019ApJ...880...65Y} {880, 65}

\bibitem[\protect\citeauthoryear{{Yencho}, {Johnston}, {Bullock}  \&
  {Rhode}}{{Yencho} et~al.}{2006}]{yencho06}
{Yencho} B.~M.,  {Johnston} K.~V.,  {Bullock} J.~S.,   {Rhode} K.~L.,  2006,
  \mn@doi [\apj] {10.1086/502619}, \href
  {https://ui.adsabs.harvard.edu/abs/2006ApJ...643..154Y} {643, 154}

\bibitem[\protect\citeauthoryear{{de Jong} et~al.,}{{de Jong}
  et~al.}{2019}]{4most}
{de Jong} R.~S.,  et~al., 2019, \mn@doi [The Messenger]
  {10.18727/0722-6691/5117}, \href
  {https://ui.adsabs.harvard.edu/abs/2019Msngr.175....3D} {175, 3}

\makeatother
\end{thebibliography}

\label{lastpage}
\end{document}